\documentclass[hyper,a4paper]{JHEP3}
\usepackage{multirow}
\pdfoutput=1
\usepackage{subfigure}
\usepackage{rotating}
\usepackage{amssymb}
\usepackage{amsmath}

\newcommand{\bea}{\begin{equation}}
\newcommand{\eea}{\end{equation}}
\newcommand{\be}{\begin{eqnarray}}
\newcommand{\ee}{\end{eqnarray}}

\def\hbar#1{\backslash\hspace{-2mm}#1}

\catcode`\@=11
\def\lsim{\mathrel{\mathpalette\@versim<}}
\def\gsim{\mathrel{\mathpalette\@versim>}}
\def\@versim#1#2{\vcenter{\offinterlineskip
\ialign{$\m@th#1\hfil##\hfil$\crcr#2\crcr\sim\crcr } }}
\catcode`\@=12

\def\2tvec#1#2{
\left(
\begin{array}{c}
#1  \\
#2  \\
\end{array}
\right)}

\def\mat2#1#2#3#4{
\left(
\begin{array}{cc}
#1 & #2 \\
#3 & #4 \\
\end{array}
\right) }

\def\Mat3#1#2#3#4#5#6#7#8#9{
\left(
\begin{array}{ccc}
#1 & #2 & #3 \\
#4 & #5 & #6 \\
#7 & #8 & #9 \\
\end{array}
\right) }

\def\3tvec#1#2#3{
\left(
\begin{array}{c}
#1  \\
#2  \\
#3  \\
\end{array}
\right)}

\def\to{\rightarrow}

\def\hbar#1{\backslash\hspace{-2mm}#1}

\numberwithin{equation}{section}

\title{Status of $Y=0$ Triplet Higgs with supersymmetry in the light of $\sim 125$ GeV  Higgs discovery}

\author{Priyotosh Bandyopadhyay$^{a,1}$,  Katri Huitu$^{a,2}$ and Asl{\i} Sabanc{\i}$^{a,3}$\\

$^a$Department of Physics,  and Helsinki Institute of Physics,\\
P.O.Box 64 (Gustaf H\"allstr\"omin katu 2), FIN-00014 University of Helsinki, Finland\\
Email: \email{$^1$priyotosh.bandyopadhyay@helsinki.fi, $^2$katri.huitu@helsinki.fi,
$^3$asli.sabanci@helsinki.fi}}

\abstract{We study the $Y=0$ triplet extended supersymmetric model in the light of the recent Higgs boson discovery. 
We calculate full one loop Higgs mass spectrum in this model 
where the possible doublet-triplet mixing is considered in the charged Higgs sector. This mixing changes the prediction of $\mathcal{B}r(B_s\to X_s \gamma)$ in this model, compared to the MSSM. The constraints from the  $\sim 125$ GeV Higgs  along with  $\mathcal{B}r(B_s\to X_s \gamma)$ are incorporated to find out the allowed parameter space. The lower bounds on the third generation squark masses coming from 125 GeV Higgs are rather week in this scenario compared to most constrained supersymmetric scenarios, e.g. a 200 GeV squark mass is still possible. }

\preprint{HIP-2013-09/TH}

\begin{document}
\section{Introduction}\label{sec:intro}
The discovery of a Higgs boson with a mass around 125 GeV has been reported by the CMS and ATLAS collaborations \cite{Higgsd1,Higgsd2}. The recent updates on the Higgs search show that $\sim 6.6 \sigma$ excess is observed in $h\rightarrow ZZ$ decay \cite{ZZexcess} and the updated mass of the Higgs boson is measured to be $125.5 \pm 0.2 {(\rm{stat})}^{+0.5}_{-0.6} (\rm{sys})$ GeV and $125.7 \pm 0.3 (\rm{stat}) \pm 0.3 (\rm{sys})$ GeV at the ATLAS and CMS, respectively \cite{newmasses}. 
In addition to the $ZZ$ decay channel, also Higgs decay to two photons gives accurate mass information.
The signal significance of the $H\to \gamma \gamma$ mode is $3.2-3.9\sigma$ in case of CMS \cite{cms2g} whereas for ATLAS it is around 6.1-7.4$\sigma$ \cite{atlas2g}. 
The Higgs boson decays to the $W$ and $Z$ pairs along with the $\tau$ and $b$ pairs are still below $5\sigma$ reach.
The spin studies in the different Higgs decay channels show that the data strongly favor the spin 0 possibility and exclude the spin 2 hypothesis with a confidence level above $99.9\%$. However, further studies and data are needed to determine whether the observed spin 0 particle is  the Standard Model (SM) Higgs or a non-standard Higgs. Non-standard Higgs structure arises in many extensions of the SM motivated {\it e.g.} by the naturalness of the Higgs mass.

Supersymmetric models remain among the best motivated extensions.
The constrained minimal supersymmetric model (CMSSM) \cite{cMSSM}, includes minimal set of parameters and has a dark matter candidate. 
Recent studies \cite{cmssmstd} 
show that it is difficult to generate a Higgs boson with mass around 125 GeV consistent with all experimental constraints, in particular dark matter relic abundance and muon $g-2$  in mSUGRA/CMSSM. 
In the general MSSM this can be achieved, if mixing between the two stops with masses less than TeV is large or if there is no mixing with very heavy stops, see {\it e.g.} \cite{mssmsd}. 
In the large mixing scenario very high $\tan{\beta}$ and $A_t$ of the order of a few TeV are needed. 
In a study with the recent squark and gluino mass limit from LHC, the possibility of 125 GeV Higgs in various supersymmetric models has been discussed in \cite{naturalsusy}.


It is well known that any low-energy supersymmetric model must be broken softly in order not to regenerate the quadratic divergences appearing in the SM Higgs sector. 
In a softly broken supersymmetry (SUSY) theory one introduces 
soft breaking parameters that may contain complex phases for explicit CP violation. 
New source of CP violation is crucial since the amount predicted by the SM through the CKM mechanism is not 
enough to explain the baryon asymmetry of the universe \cite{baryon}. 
However, it was shown in Ref.\cite{EDMs} that soft SUSY breaking CP phases generally lead to very large electric dipole 
moments (EDMs). 
An unnatural solution to this problem is fine-tuning of parameters that appear in the different sectors of the model. On the other hand,  overproduction of CP violation can also be avoided in supersymmetric theories via introducing new singlet or triplet superfield(s) whose scalar component breaks the CP symmetry spontaneously \cite{SCPVsinglet,SCPVtriplet}. Spontaneous CP violation (SCPV) is an attractive method since the CP phases are introduced only in the vacuum expectation values (vevs) of neutral Higgs fields so that the number of free CP phases is reduced naturally. It is well known that SCPV is not possible in the MSSM even with the radiative corrections because of the lightest Higgs mass bound \cite{SCPVMSSM}. This is 
why the extended MSSM Higgs sector is indispensable for spontaneously broken CP symmetry. Apart from SCPV, the triplet extended Higgs sector can also generate additional contributions to the lightest Higgs mass so that the excluded parameter region of the third generation squark masses and other soft parameters can be reopened. 

An extended supersymmetric standard model containing a triplet with hypercharge $Y=0$ or $Y=\pm2$ consists of new neutral Higgs boson(s) in addition 
to the ones from two Higgs doublets. Introduction of new neutral element with non-zero vev breaks the custodial $SU(2)_c$ symmetry of the Higgs sector 
when the gauge symmetry is broken. 
As a consequence of $SU(2)_c$ breaking,  non-zero triplet vev contributes to the tree-level prediction of the electroweak $\rho$ parameter,
\bea
\rho=\frac{\sum \limits_{T,Y}[4T(T+1)-Y^2]|v_{T,Y}|^2 c_{T,Y}}{\sum \limits_{T,Y} 2Y^2 |v_{T,Y}|^2},
\label{rho}
\eea
which in the SM is exactly one.
Here $v_{T,Y}$ are the vev's of the scalar fields with the third component of the isospin $T$ and hypercharge $Y$ whereas $c_{T,Y}=1\,(1/2)$ is for the complex (real) representation of the scalar fields. The large tree-level deviations from unity can be avoided by two ways: (i) the neutral triplet fields can have much smaller vacuum expectation values (vevs) than those for the 
neutral doublet fields; or (ii) 
the triplet fields and the vevs of their neutral members can be arranged in such a way that the custodial $SU(2)_c$ symmetry is preserved \cite{tripletlit}. 
We follow the former path where a non zero vev is generated for the neutral triplet field  which is strongly constrained by the global fit on $\rho$ parameter 
measurements \cite{PDG}. 

In this work we consider the triplet extended supersymmetric standard model (TESSM) where we expand the field content 
of MSSM by adding a new SU(2) triplet chiral superfield with hypercharge $Y=0$.
The TESSM has been extensively studied in Ref.~\cite{triplet} where all tree level stability conditions of the scalar potential as well as the 
Higgs mass spectrum have been calculated. It is well known that the new  $Y=0$ triplet superfield generates additional radiative corrections to the lightest Higgs mass. The calculation of Higgs masses at one loop level have been addressed in Ref.~\cite{rho,chargedH} in some specific contexts. 
In Ref.~\cite{rho} the electroweak sector contributions have been considered and it is found that the Higgs boson mass can be up to $140$ GeV. In Ref.~\cite{chargedH} the strong and Higgs sector contributions are considered when the D-terms are omitted. For a complete analysis one should calculate all possible radiative corrections coming from both the strong and the electroweak sectors where the latter contributions are negligible in the MSSM. Such analysis may reopen some of the regions of the third generation sfermion masses and other soft parameters which are excluded in the MSSM like models in the context of recent observation of $\sim 125$ GeV Higgs.  
TESSM has also been studied in \cite{Delgado:2012sm,Delgado:2013zfa} where in \cite{Delgado:2012sm} especially the diphoton
decay rate of the lightest Higgs and in \cite{Delgado:2013zfa} also other production and decay channels were considered.
It was found that the current experimental information can be satisfied, whether the diphoton rate is enhanced or not.
In \cite{Delgado:2012sm,Delgado:2013zfa} the triplet vev has been ignored and in the Higgs sector only the strong sector corrections 
and a very heavy triplet are included. In our work we consider the parameter space of the model in more detail than is previously done by 
including full one-loop corrections, and indeed find that including both strong and electroweak sector is important.
We constrain the parameter space by taking into account experimental mass limits and the decay $B_s \to X_s \gamma$.

We organize our paper as follows. In Section~\ref{THM} we review the relevant features of the model we use and discuss the Higgs spectrum of the triplet extended supersymmetric model. 
We consider 
the radiative corrections to the neutral Higgs masses from both the strong and the electroweak sectors. In Section~\ref{param} we study the 
parameter space in detail. 
We show that it is possible to have a $\sim125$ GeV Higgs in the spectrum without introducing any large mixing in the sfermion sector or very heavy third generation sfermion masses. Section~\ref{bsg} is devoted to the $B_s \to X_s \gamma$ analysis for the TESSM model where the lack of fermion-triplet coupling can alter the contributions to  $B_s \to X_s \gamma$ in comparison with the ones in the MSSM models. We show that the triplet component of the charged Higgs boson has a significant impact on the $B_s \to X_s \gamma$ calculation. The triplet nature reduces the charged Higgs boson contributions to the  $\mathcal{B}r(B_s \to X_s \gamma)$ such a way that the total contribution can be enhanced or suppressed depending on the relative sign of $\mu_D A_t$ in the chargino contributions. In Section~\ref{allowedbsg} we combine the constraints on the parameter space coming from the Higgs discovery with the ones from the $B_s \to X_s \gamma$ analysis. We observe that it is possible to obtain many parameter regions with a $\sim125$ GeV Higgs that satisfy the experimental value of $\mathcal{B}r(B_s \to X_s \gamma)$ within $\pm 2 \sigma
 $. Finally we conclude in Section~\ref{conclu}.

\section{$Y=0$ triplet extended Higgs sector}\label{THM}

In TESSM, in addition to the MSSM like two Higgs doublets, there is an $SU(2)$ complex Higgs triplet with zero hypercharge which 
can be represented as a 2x2 matrix

 \begin{equation}
 {\bf \Sigma} = \begin{pmatrix}
       \sqrt{\frac{1}{2}}\xi^0 & \xi_2^+ \cr
      \xi_1^- & -\sqrt{\frac{1}{2}}\xi^0
       \end{pmatrix} .
 \end{equation}
Here $\xi^0$ is a complex neutral field, while  $\xi_1^-$ and $\xi_2^+$ are the charged Higgs fields. 
Note that $(\xi_1^-)^*\neq -\xi_2^+$. The triplet field $\Sigma$ couples to the two Higgs doublets by a dimensionless 
coupling $\lambda$, which is a free parameter of the model \cite{triplet}. 
Thus the superpotential of  the Higgs sector of the model
is given by

\begin{equation}\label{superpot}
W=\lambda H_d.\Sigma H_u\, +\, \mu_D H_d.H_u \, +\, \mu_T Tr(\Sigma^2),
 \end{equation}
where $\mu_D$ is the usual mixing parameter of the two MSSM like Higgses and $\mu_T$ is the mass parameter of the triplet. The Higgs potential can be 
calculated by collecting relevant terms from superpotential in Eqn.~(\ref{superpot}), soft breaking terms and the D-terms as follows

\begin{equation}\label{Higgspot}
 V=V_F \,+V_D\,+\, V_S,
\end{equation}
where $V_F$ and $V_D$ are contributions from the F-terms and D-terms, which can be found in Appendix. $V_S$ contains the 
soft-supersymmetry breaking terms:
\be
V_S&= &m_1^2 |H_d|^2 + m_2^2 |H_u|^2 + m_3^2 {\rm Tr} (\Sigma^\dag \Sigma)\\\nonumber
& +  &[A_{\lambda} \lambda H_d \Sigma H_u + B_D \mu_D H_d  H_u + B_T \mu_T {\rm Tr} (\Sigma^2) + {\rm H.c.} ] .\
\ee
Here $A_{\lambda}$ is the soft trilinear parameter, $B_D$ and $B_T$ are the soft bilinear parameters while $m_i$ ($i=1,2,3$) represent the soft SUSY breaking masses. 
For simplicity we assume here that there is no CP violation in the Higgs sector so that all the parameters as well as the vacuum expectation 
values of the neutral Higgs fields are chosen to be real. 
When these neutral fields acquire non-zero vevs, the electro-weak symmetry (EWS) is spontaneously 
broken and all fermions and gauge bosons gain masses.
We denote
\be
\langle H_u^0 \rangle = v_u,\;\; \langle H_d^0 \rangle = v_d,\;\; \langle \xi^0 \rangle = v_T, 
\ee
and $\tan{\beta}={v_u}/{v_d}$.
The W boson mass expression is altered by the triplet vev as 
$m_W^2=g_2^2(v^2+4v_T^2)/2$, where $v^2=v_u^2+v_d^2$; whereas the Z boson mass expression remains unaffected. 
As seen from Eqn.~(\ref{rho}),
\bea
\rho=1+4v_T^2/v^2.
\eea
Thus the triplet vev is strongly constrained by the global fit on the $\rho$ 
parameter measurement \cite{PDG},
\bea
\rho=1.0004{\small{\begin{array}{l} + 0.0003\\-0.0004 \end{array} }}
\eea
which implies $v_T\leqslant3$ GeV. 
We have used  $v_T=3$ GeV in our numerical analysis. 

The scalar Higgs potential Eqn.~(\ref{Higgspot}) can be split up into charged and neutral parts for Higgs sector analysis. The three minimization conditions obtained from the neutral  part allow us to write the Higgs sector soft masses ($m_i$, i=1,2,3) in terms of the other parameters of the model. Tree level expressions of the corresponding soft masses and the squared mass matrices of the neutral Higgs sector as well as the charged Higgs sector can be found in Ref.\cite{triplet}. After the electroweak symmetry breaking, the physical Higgs bosons of the TESSM comprise three CP-even ($h$, $H_{1,2}$), two CP-odd ($A_{1,2}$) and three charged Higgs bosons ($H^{\pm}_{1,2,3}$). In this context $h$ corresponds to the lightest Higgs boson of the model, whereas the others are generally much heavier. It was shown in Ref.~\cite{rho} that it is possible to obtain the lightest Higgs boson with a mass up to $120$ GeV at tree level but for a $~125$ GeV Higgs one also needs to consider the radiative corrections to the neutral Higgs sector. The one-loop radiative corrections to the Higgs potential can be calculated using the effective potential approach \cite{coleman-weinberg}
\bea
\Delta V=\frac{1}{64\pi^2}Str\left[\mathcal{M}^4\left(ln\frac{\mathcal{M}^2}{\Lambda^2}-\frac{3}{2}\right)\right].
\eea
Here $\mathcal{M}$ represents the field dependent mass matrices of the particles and $\Lambda$ is the renormalization scale. As the scalar potential changes with the one loop corrections one needs to recalculate the soft masses which depend also on the particles circulating in the loop.  

So far in the literature there is no calculation of full one loop contribution from all sectors. In Ref.~\cite{rho} authors have shown that for large $\lambda \sim 0.8-0.9$, the electroweak contributions are sufficient 
to raise the Higgs boson mass to 140 GeV whereas the gaugino-Higgsino mixing was ignored. 
In Ref.~\cite{chargedH} the authors considered only the strong and Higgs 
sectors in their one loop calculation of the Higgs masses. However the analysis has been done only for a specific choice of parameter space, where D-terms are omitted and it is assumed that Higgs-higgsino loops only modify the $(3,3)$ elements of the CP even Higgs mass matrix of Ref.~\cite{chargedH} for large $\lambda$. 
In the current study, we consider the radiative corrections from both the strong and electro-weak sectors.  The one-loop effects from these sectors for each entry in the Higgs mass matrix have been incorporated. It is known that the dominant radiative corrections to the lightest Higgs mass come from the top-stop loops because of the large Yukawa couplings. Unlike low $\tan{\beta}$ case, for large $\tan\beta$, one should also consider the bottom-sbottom loops as the corrections are non-negligible. We discuss in Section \ref{param}, that  even though the dominant radiative corrections are obtained from the strong sector, electroweak sector still has an important effect on Higgs masses emerging both in the minimization conditions and the mass matrix entries.


\section{Higgs sector with $m_h \sim 125$ GeV}\label{param}

In this section we show how the different sectors of the TESSM can affect the Higgs self-energies at one loop.
For this purpose we consider both the strong sector and the electroweak sector. In MSSM 
the most significant contributions to the Higgs boson masses can be evaluated using the diagrams containing top and stop, and 
for large $\tan\beta$ also  from bottom-sbottom loop, which correction becomes comparable with the top-stop one. 
In TESSM these radiative corrections 
remain the dominant ones but for highly coupled ($\lambda$ large) TESSM the electroweak sector contributions become non-negligible.

To scan the parameter space we select four different scenarios as given in Table~\ref{Scs}. 
In these scenarios large mass differences between the third generation squarks or very heavy particles are not
needed in order to satisfy the experimental constraints.
We fix the third generation squark masses  and the triplet soft mass $\mu_T$ whereas the supersymmetric bilinear mass term $\mu_D$ and triplet Higgs coupling $\lambda$ are free parameters. 
 In this analysis we use several values of $\tan\beta$ for each scenario and top mass is taken as 173.2 GeV \footnote{Current value of top quark mass is $173.18 \pm 0.56 (\rm{stat}) \pm 0.75 (\rm{syst})$ GeV or $173.18 \pm 0.94$ GeV \cite{mtop}.}.
\begin{table}
\begin{center}
\renewcommand{\arraystretch}{1.4}
\begin{tabular}{||c|c|c|c||}
\hline\hline
Scenario&$m_{\tilde{t}_1,\tilde{b}_1}$&$m_{\tilde{t}_2,\tilde{b}_2}$&$\mu_T$\\
  &(GeV)&(GeV)&(GeV)\\

\hline\hline
Sc1 &500 &550 & 500\\
\hline
Sc2 &500 &550 & 1200\\
\hline
Sc3 &1000 &1050 & 500\\
\hline
Sc4 &1000 &1050 & 1200\\
\hline
\hline
\end{tabular}
\caption{Scenarios for the allowed parameter space.}\label{Scs}
\end{center}
\label{bmp}
\end{table}
 
\begin{figure}
\centering
\mbox{\subfigure[]{\includegraphics[width=3in,height=2.5in]{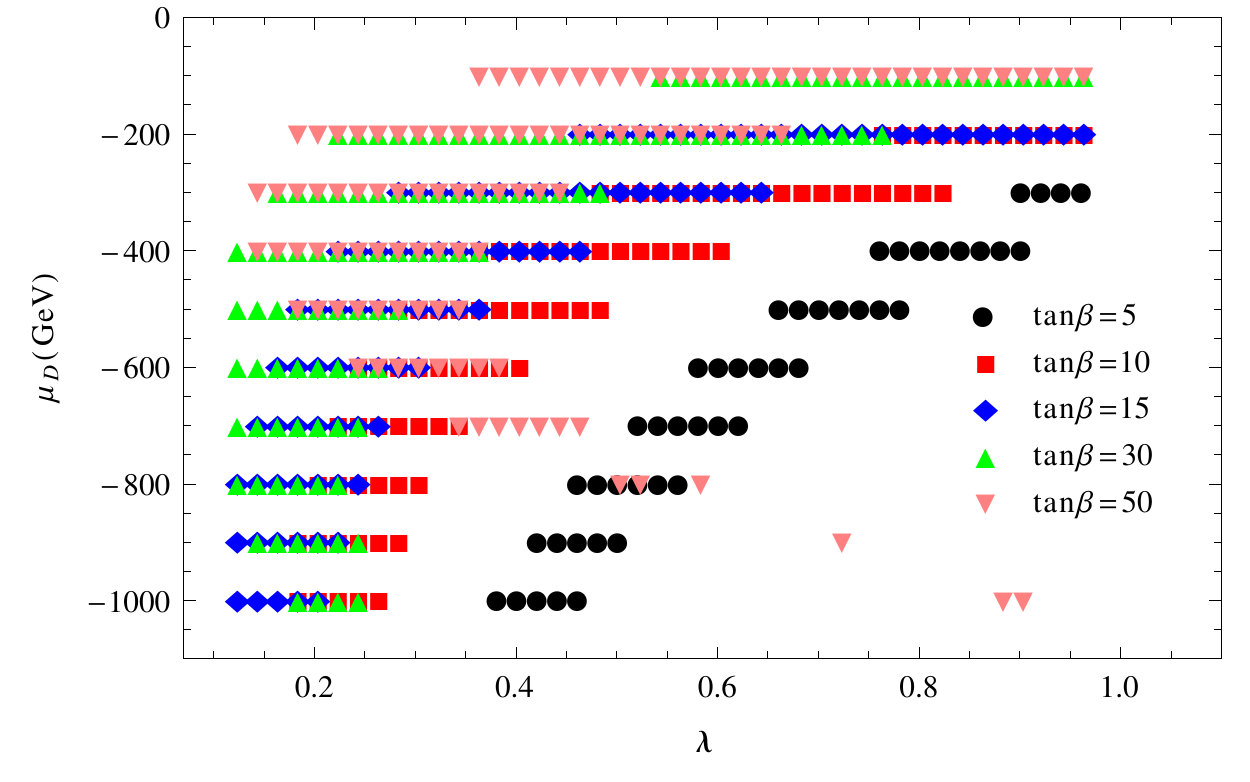}}\quad
\subfigure[]{\includegraphics[width=3in,height=2.5in]{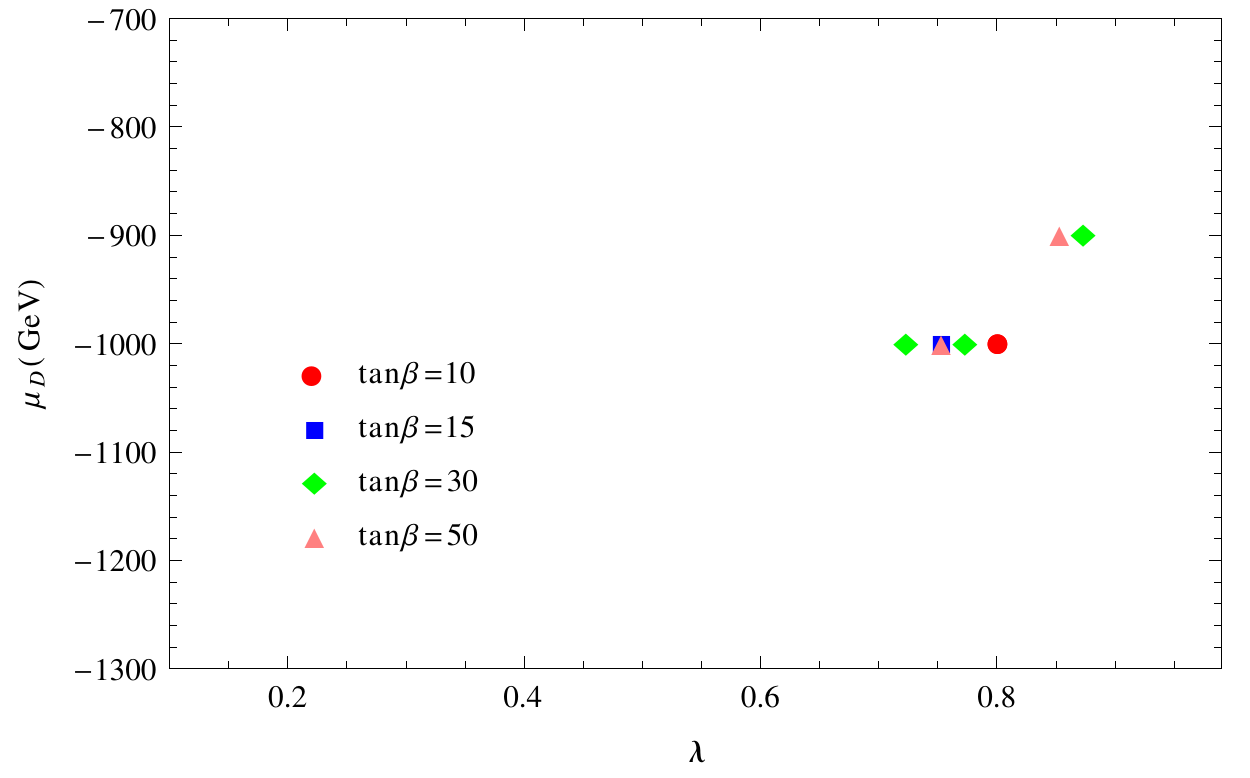} }}
\quad
\subfigure[]{\includegraphics[width=3in,height=2.5in]{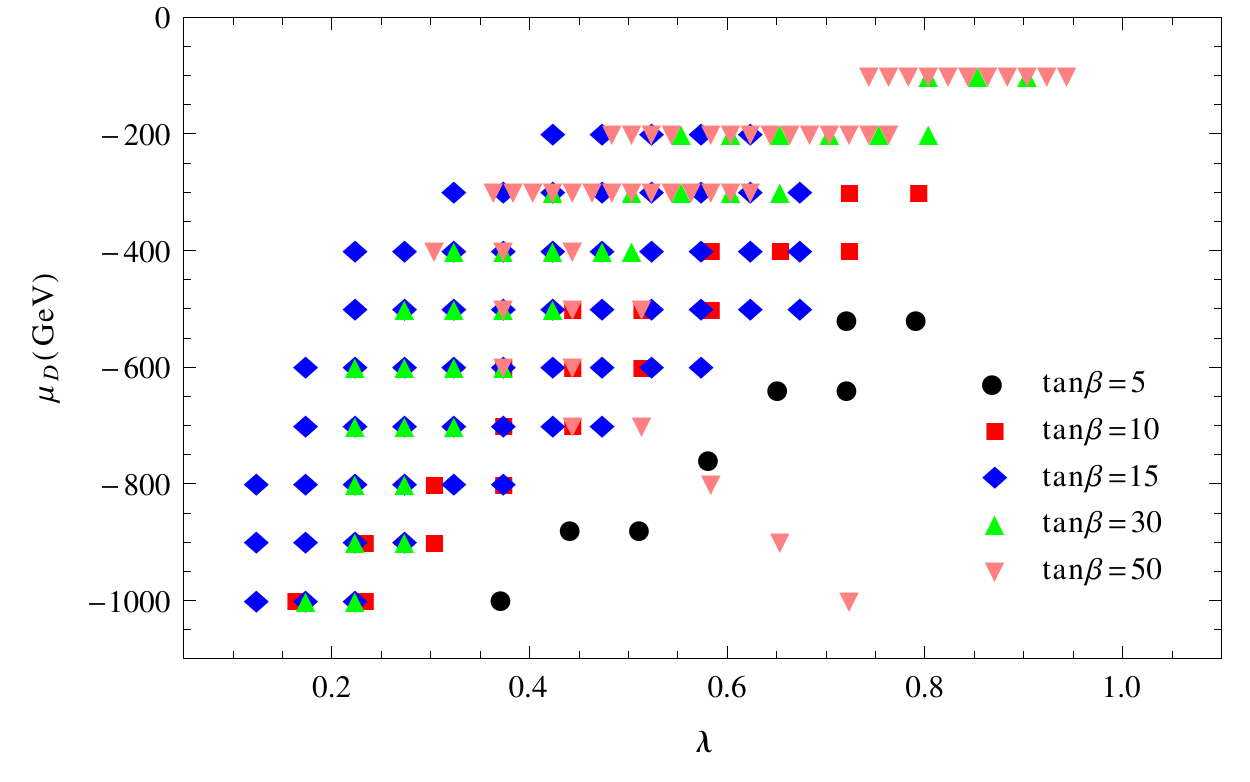} }
\caption{Allowed parameter space in $\mu_D-\lambda$ plane for the scenario 1 (Sc1) where (a) only strong sector (b) only weak sector and (c) total contributions are taken into account.}\label{sc1}

\end{figure}

 In Fig.~\ref{sc1} we show the allowed region of the parameter space for $\mu_D-\lambda$ plane where the lightest Higgs mass remains within 124-127 GeV for 
Sc1 of Table~\ref{Scs}. 
It is seen that the desired Higgs mass prefers the $\mu_D<0$ region for this scenario.  
In Fig.~\ref{sc1}(a) radiative corrections include only contributions from the strong sector, i.e, from top-stop and bottom-sbottom as 
explained above. 
It is clear from Fig.~\ref{sc1}(a) that for large values of $\tan{\beta}=30,50$, solutions are possible with small $\lambda$ which corresponds to the triplet sector being weakly coupled to MSSM. For smaller values of $\tan{\beta}\sim 5$, relatively larger values of $\lambda \gsim 0.4$ are needed to achieve the Higgs mass around 125 GeV. This can be understood from the nondiagonal terms in the stop and sbottom mass matrices,
\be
m_{X_t}^2 & =&  m_t\left(A_t +\mu_D \cot\beta -\frac{\lambda}{\sqrt{2}}v_T \cot\beta \right)\label{stopm},\nonumber\\
m_{X_b}^2 & = & m_b\left(A_b +\mu_D \tan\beta-\frac{\lambda}{\sqrt{2}}v_T \tan\beta \right). 
\ee
For small values of $\tan\beta$, the top-stop contributions to the Higgs mass are important where the mixing in the stop sector is large because of the top mass. Thus non-negligible triplet contribution to the mixing can be achieved as long as $\lambda$ is large. These triplet contributions can enhance the mixings to raise  the Higgs mass to $\sim125$ GeV without having large values of $A_q$ and $\mu_D$. For large values of $\tan\beta$ the contribution to the Higgs mass from the sbottom sector becomes significant, when the mixing in the sbottom sector is large. The mixing terms proportional to $\lambda$  and $\mu_D$ increase with $\tan{\beta}$ and when these terms get additive, the bottom mixing is further enhanced such that small value of $\lambda$ is enough to obtain the Higgs mass around 125 GeV. Thus this triplet contribution to the mixing can be sufficient to achieve $\sim 125$ GeV Higgs even in the case of the exact cancellations between the $A_b$ and $\mu_D$ terms.  

In Fig.~\ref{sc1}(b) radiative corrections include only the electroweak (EW) sector. We assume here that the soft gaugino masses are large enough, so that they are decoupled. We are interested to see if the higgsino and so the triplet contribution is enough to have $m_h\sim 125$ GeV, 
{\it i.e.} we ignore the
gaugino contribution and gaugino-higgsino mixing.\footnote{For a complete study one has to take into account the gaugino-higgsino mixing, which extends the neutralino and chargino sectors. We are considering the gaugino-higgsino mixing in the context of complete neutralino mass matrix in this model in \cite{PBKHAS}.} 
It is interesting to see that we do have some points where the Higgs boson mass is  $\sim 125$ GeV, even only from the electroweak corrections as was earlier reported in \cite{rho}.  These solutions prefer larger values of $\lambda$, which implies that a highly coupled triplet scenario is necessary to have large enough electro-weak contribution.

In Fig.~\ref{sc1}(c) we consider the radiative contributions coming from both the strong and EW sectors. The generic features remain the same as in Fig.~\ref{sc1}(a) where only the strong sector contributes. A close look however points to the fact that the allowed region of parameter space is pushed to smaller values in $\mu_D$ and larger
values in $\lambda$ compared to Fig.~\ref{sc1}(a). 
This happens since the minimization conditions change, when both contributions are included.
Therefore, the combination of both sectors demonstrates the importance of including all the relevant contributions.

\begin{figure}
\centering
\mbox{\subfigure[]{\includegraphics[width=3in,height=2.5in]{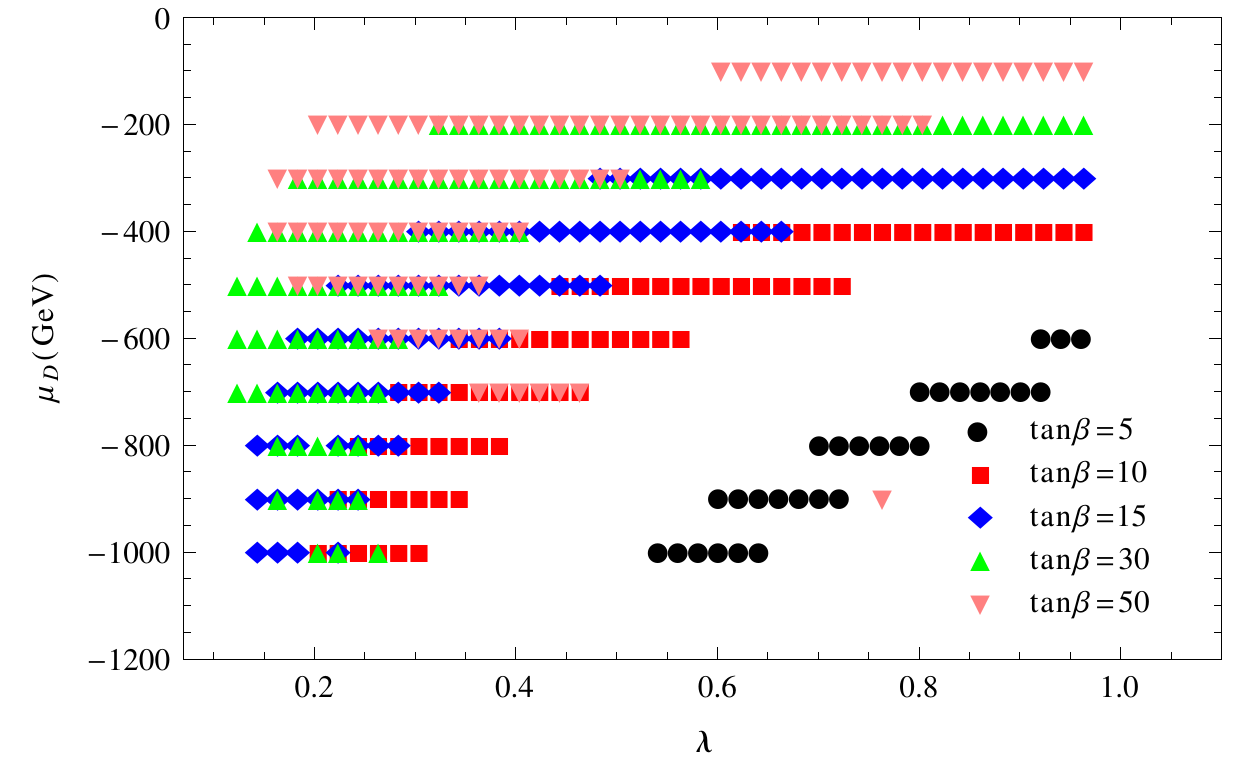}}\quad
\subfigure[]{\includegraphics[width=3in,height=2.5in]{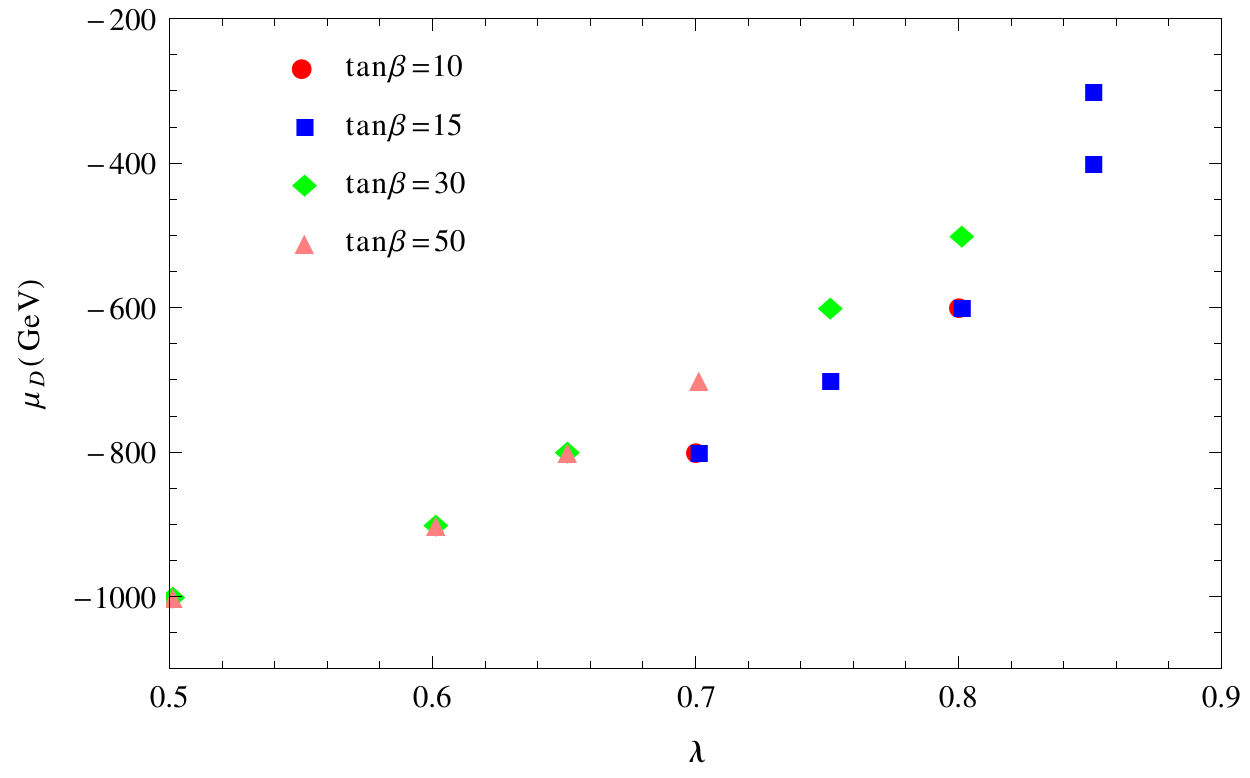} }}
\quad
\subfigure[]{\includegraphics[width=3in,height=2.5in]{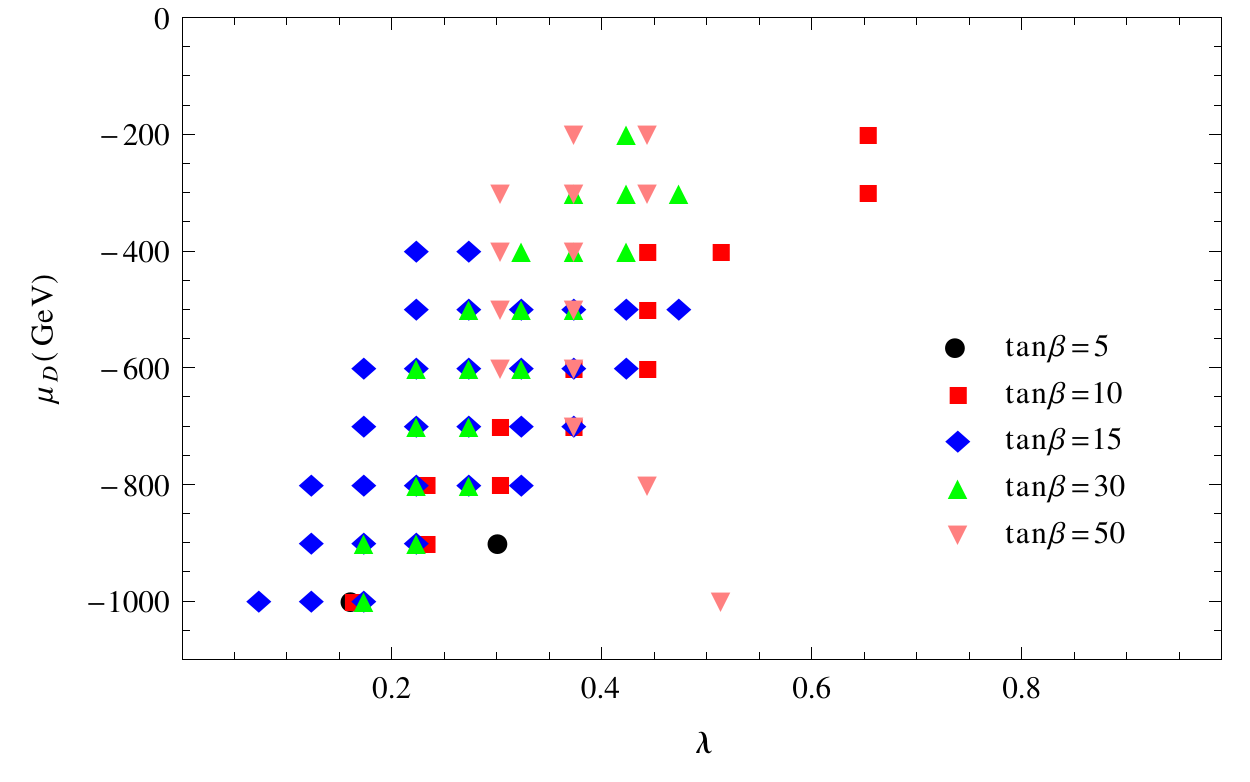} }
\caption{Allowed parameter space in $\mu_D-\lambda$ plane for the scenario 2 (Sc2) with different $\tan{\beta}$ where (a) only strong sector, (b) only weak sector and (c) total contributions are considered.}\label{sc2}
\end{figure}

In Fig.~\ref{sc2} the allowed region in $\mu_D-\lambda$ plane is plotted for 
scenario 2 (Sc2) in which $\mu_T$ is raised to 1.2 TeV for various $\tan{\beta}$ values.
It is seen that for small $\tan{\beta}$ ($=$5), the allowed values of $\lambda$ vary between 0.6-0.9, 
whereas for large $\tan{\beta}$ also small values $\lambda(\leq 0.4)$ are allowed.  
In Fig.~\ref{sc2}(b) the case with only EW sector 
contribution at the one-loop level is shown.
Similar to Sc1, some solutions for highly coupled Higgs sector (with $\lambda\geq 0.5$) are found.
Fig.~\ref{sc2}(c) 
presents the allowed regions when both the strong and EW sectors contribute at one loop level. 
As figure suggests, the change in the minimization conditions does rule out 
some of the allowed regions in Fig.~\ref{sc2}(a), but allowed parameter space exists even for $\lambda\leq 0.2$. 

\begin{figure}
\centering
\mbox{\subfigure[]{\includegraphics[width=3in,height=2.5in]{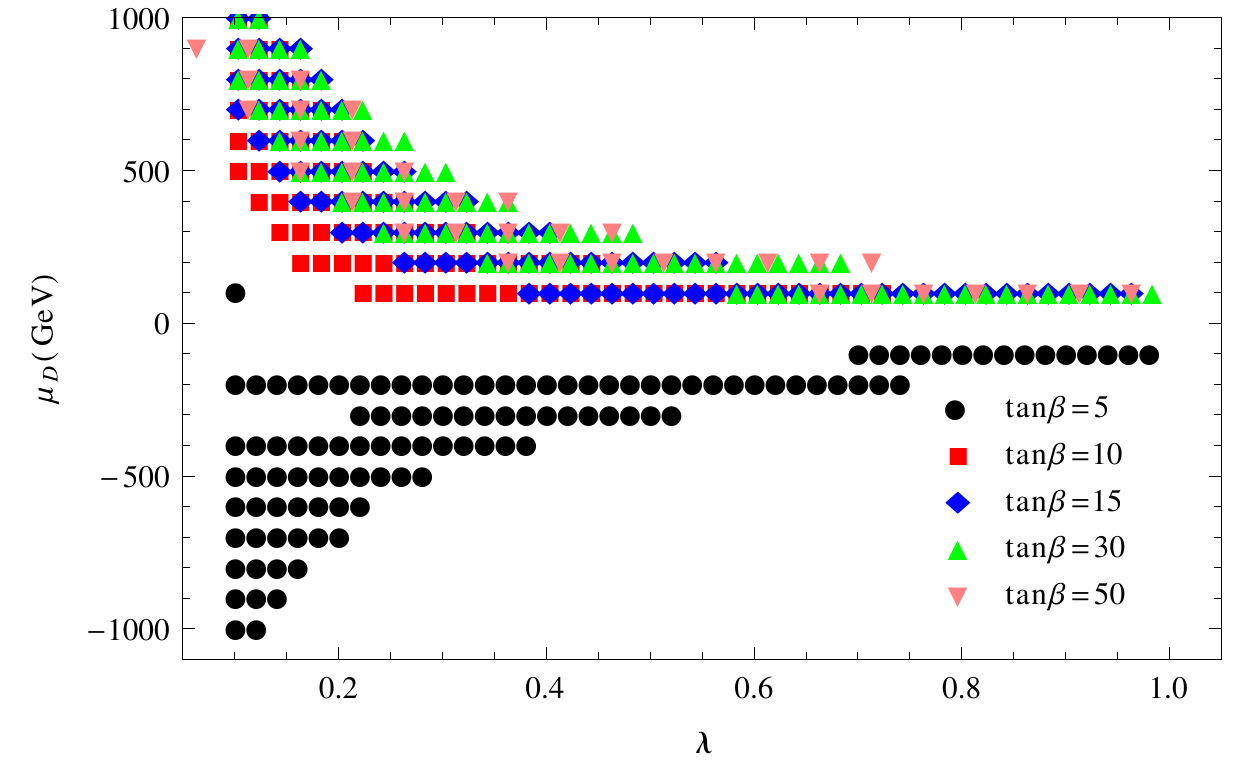}}\quad
\subfigure[]{\includegraphics[width=3in,height=2.5in]{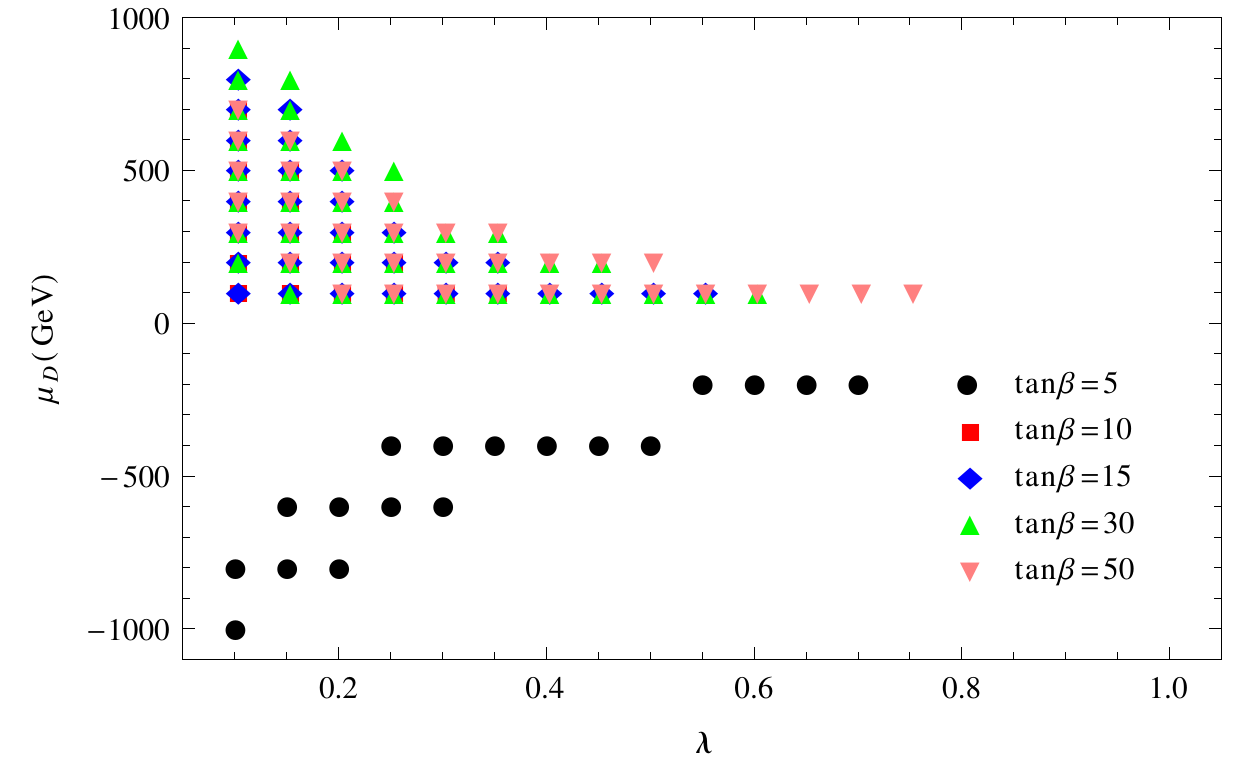} }}
\caption{Allowed parameter space in $\mu_D-\lambda$ plane for the scenario 3 (Sc3) with different $\tan{\beta}$ where (a) only strong sector
and (b) total contributions are considered.}\label{sc3}
\end{figure}

\begin{figure}
\centering
\mbox{\subfigure[]{\includegraphics[width=3in,height=2.5in]{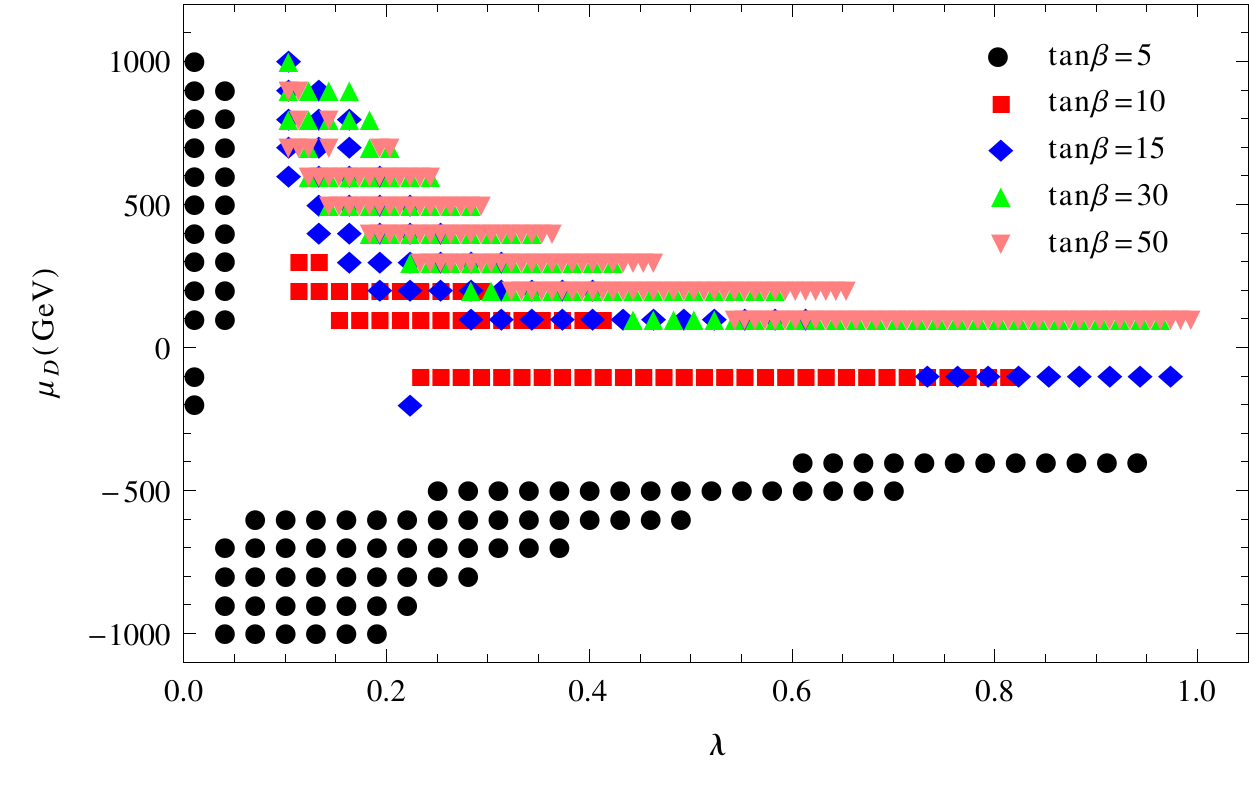}}\quad
\subfigure[]{\includegraphics[width=3in,height=2.5in]{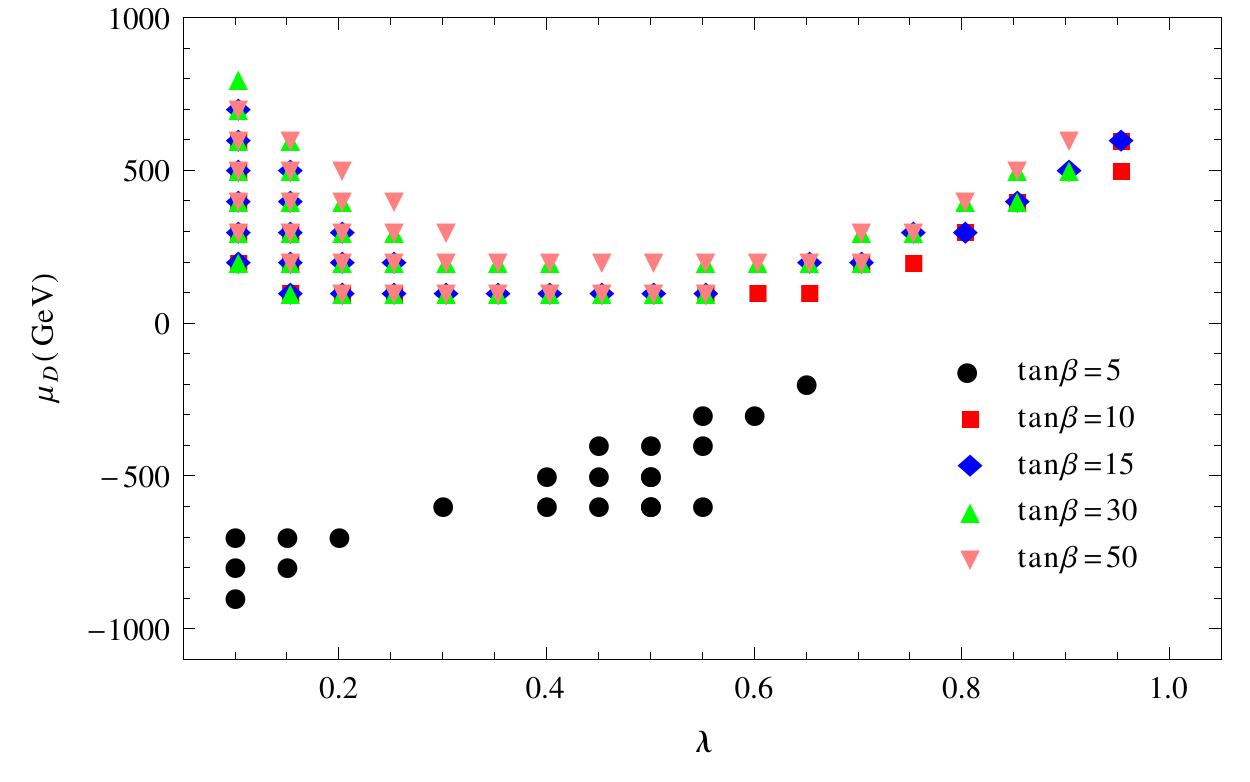} }}
\caption{Allowed parameter space in $\mu_D-\lambda$ plane for the scenario 4 (Sc4) with different $\tan{\beta}$ where (a) only strong sector, and (b) total contributions are taken into account.}\label{sc4}
\end{figure}

In Fig.~\ref{sc3} we consider the case of $m_{\tilde{t}_1}=1 $ TeV with $\mu_T=0.5$ TeV (Sc3). In Fig.~\ref{sc3}(a) it is seen that low $\tan{\beta}$ strongly favours $\mu_D < 0$ regions. For larger $\tan{\beta}$, $\mu_D > 0$ regions are allowed unlike in the previous scenarios. 
This behaviour remains even after total one-loop contributions are  
taken into account, as can be seen from  Fig.~\ref{sc3}(b). 
In scenario 4 (Sc4), where $\mu_T=1.2$ TeV as shown in  Fig.~\ref{sc4}, 
similar behaviour of $\mu_D$ changing sign as $\tan{\beta}$ increases is found.

\begin{figure}
\centering
\mbox{\subfigure[]{\includegraphics[width=3in,height=2.5in]{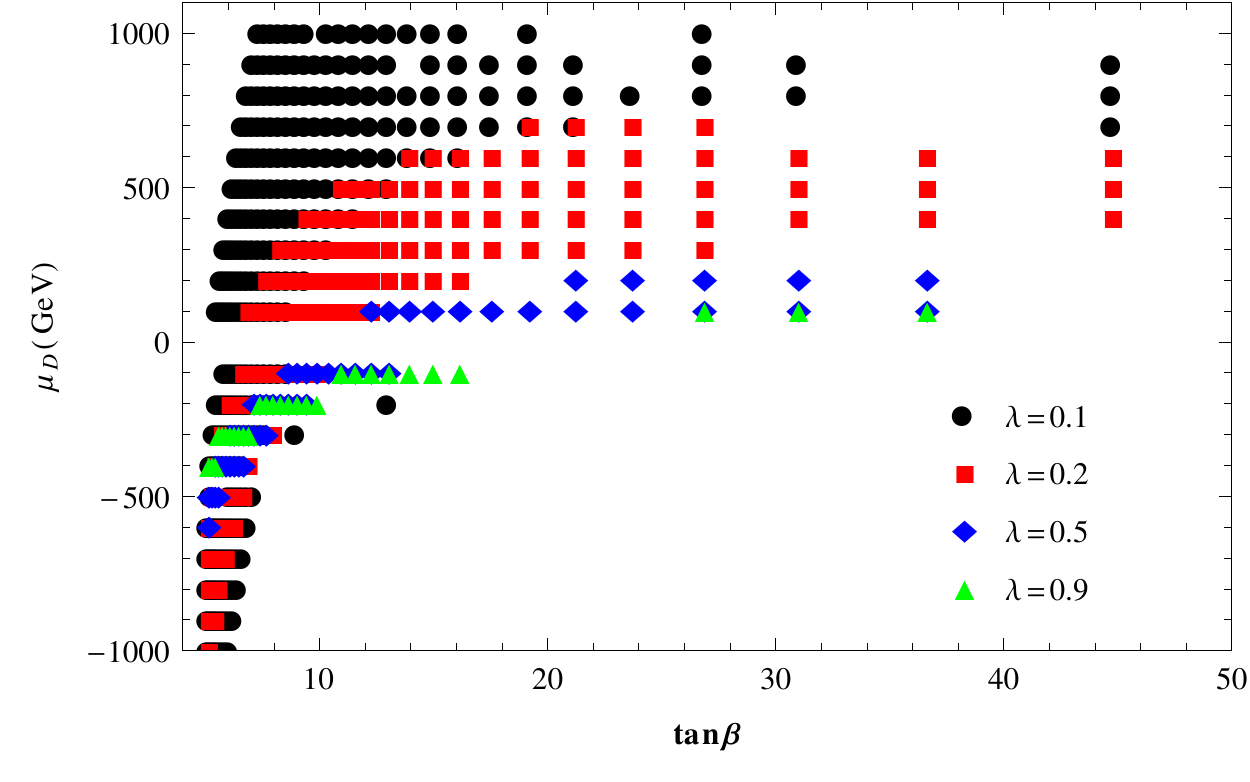}}\quad
\subfigure[]{\includegraphics[width=3in,height=2.5in]{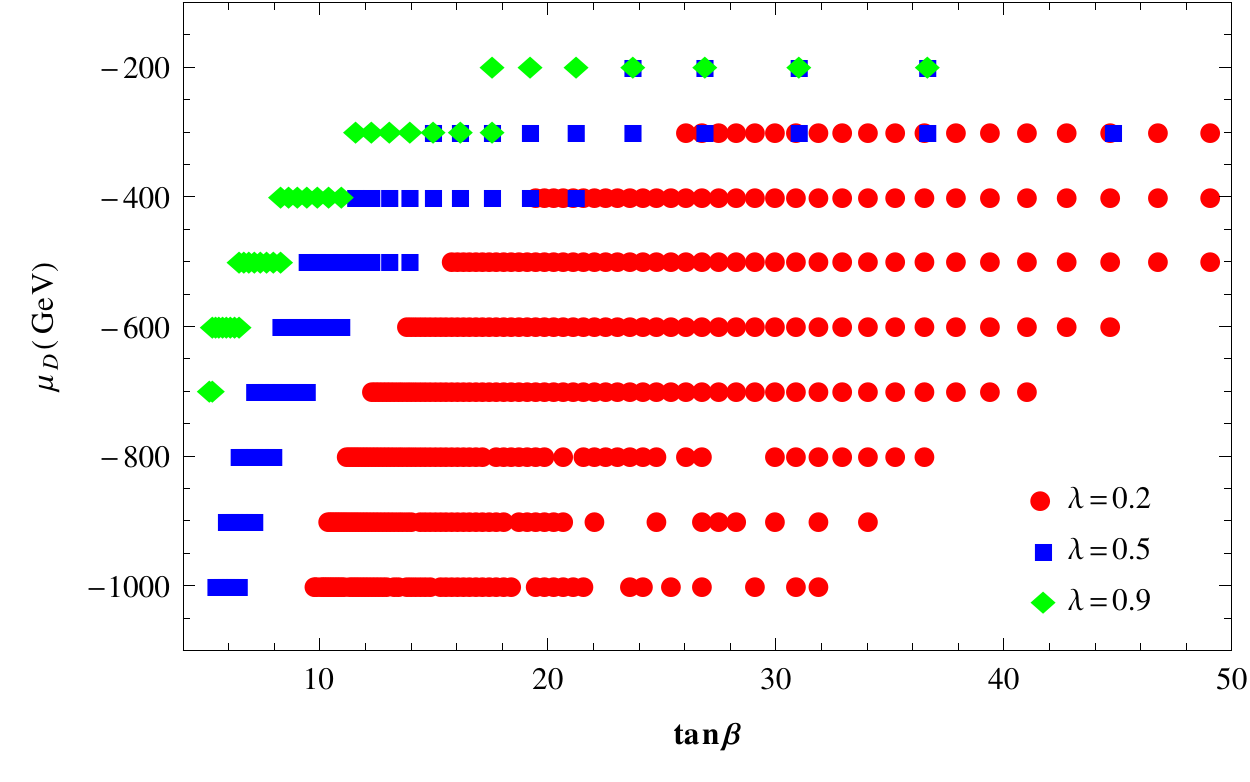} }}
\caption{Allowed parameter space in $\mu_D-\tan\beta$ plane with different $\lambda$ values for (a) scenario 4 (Sc4) and (b) scenario 2 (Sc2).}\label{tanbv}
\end{figure}

 To study the above observation in detail, we scan the allowed parameter space in $\mu_D-\tan{\beta}$ plane for different sets of $\lambda$ values in 
scenarios where  the loop contribution includes only the strong sector. 
Fig.~\ref{tanbv}(a) shows the allowed region for (Sc4), $m_{\tilde{t}_1}= 1$ TeV 
and $\mu_T = 1.2$ TeV, for  $\lambda=0.1, 0.2, 0.5, 0.9$ and Fig.~\ref{tanbv}(b) for (Sc2), $m_{\tilde{t}_1}= 500$ GeV and $\mu_T = 1.2$ TeV. 
From the Fig.~\ref{tanbv}(a) we can clearly see that for larger stop mass case ($\sim 1$ TeV), $\mu_D> 0$ is preferred for large $\tan{\beta}$ 
for both small and large values of $\lambda$. Unlike Sc4, in scenario 2 (Sc2), where we have stop and sbottom masses around 500 GeV, the allowed region 
prefers $\mu_D < 0$ for all $\tan{\beta}$ and $\lambda$ values. The interesting point to note that in Sc2 for smaller values of $\lambda$, i.e., for weakly coupled 
theory, allowed regions prefer more negative values of $\mu_D$.  This behaviour remains similar when all contributions to one-loop correction are 
taken in account.

\begin{figure}
\centering
\mbox{\subfigure[]{\includegraphics[width=3in,height=2.5in]{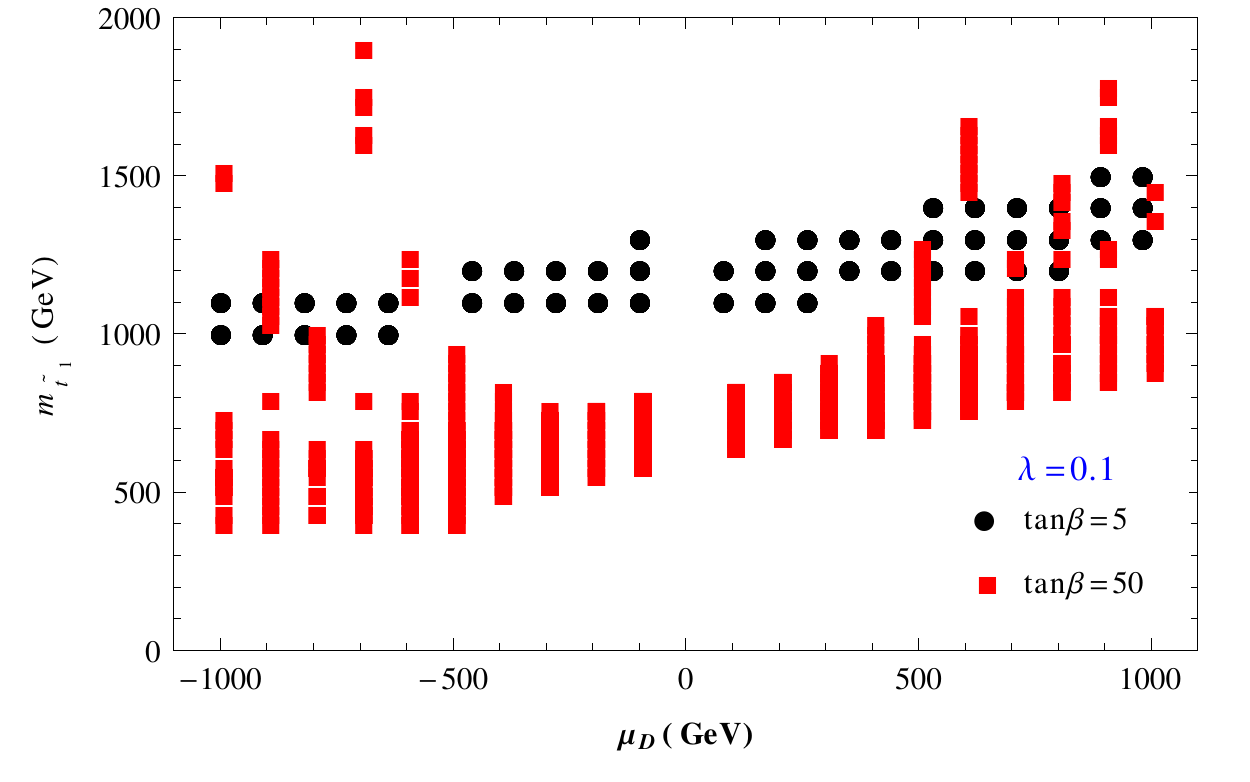}}\quad
\subfigure[]{\includegraphics[width=3in,height=2.5in]{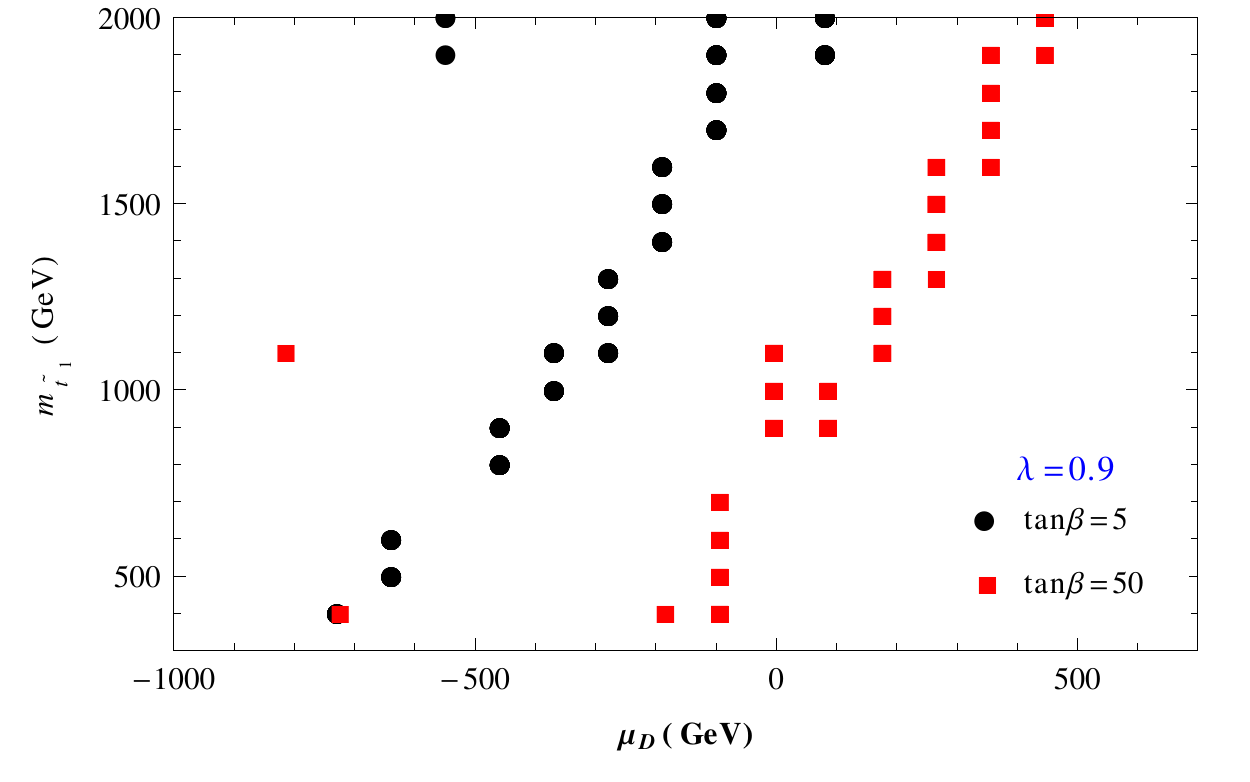} }}
\caption{The variation of  $m_{\tilde{t}_1}$ with $\mu_D$ for the lightest Higgs boson mass $\sim 125$ GeV. Two scenarios $\lambda=0.1, 0.9$ are displayed for different $\tan\beta$ values where only the strong sector contributions to the Higgs sector are taken into account.}\label{mud}
\end{figure}

To understand this behaviour, we study the stop or sbottom masses as $\mu_D$ changes sign and when the lightest Higgs mass is 
around 125 GeV. 
We plot $m_{\tilde{t}_1}-\mu_D$ for different $\tan{\beta}$ values in 
Fig.~\ref{mud} for $\lambda=0.1, 0.9$, respectively.  
Fig.~\ref{mud}(a) shows that for small $\lambda$ ($\sim 0.1$), the stop mass variation is rather small
and the $\mu_D$ values can be either positive or negative for most cases.
For the highly coupled case, from Fig.~\ref{mud}(b), it is seen that stop mass increases fast with $\mu_D$,
and large $\tan{\beta}$ prefers large values of $\mu_D$ for heavy stops.
For $\tan{\beta}=5$, $\lambda=0.1$ requires $m_{\tilde{t}_1}\gsim 1$ TeV and in general the minimum squark masses are 
larger than in the highly coupled case.


\begin{figure}
\centering
\mbox{\subfigure[]{\includegraphics[width=3in,height=2.5in]{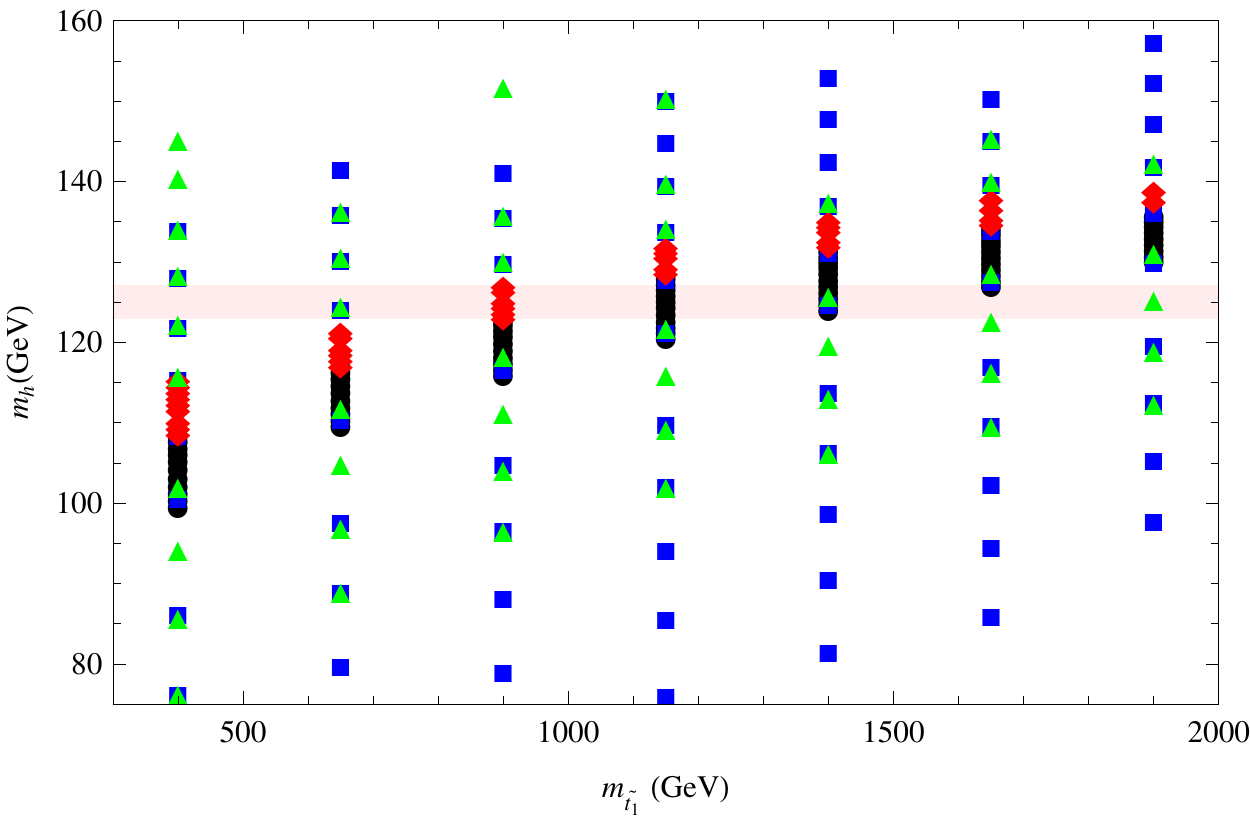}}\quad
\subfigure[]{\includegraphics[width=3in,height=2.5in]{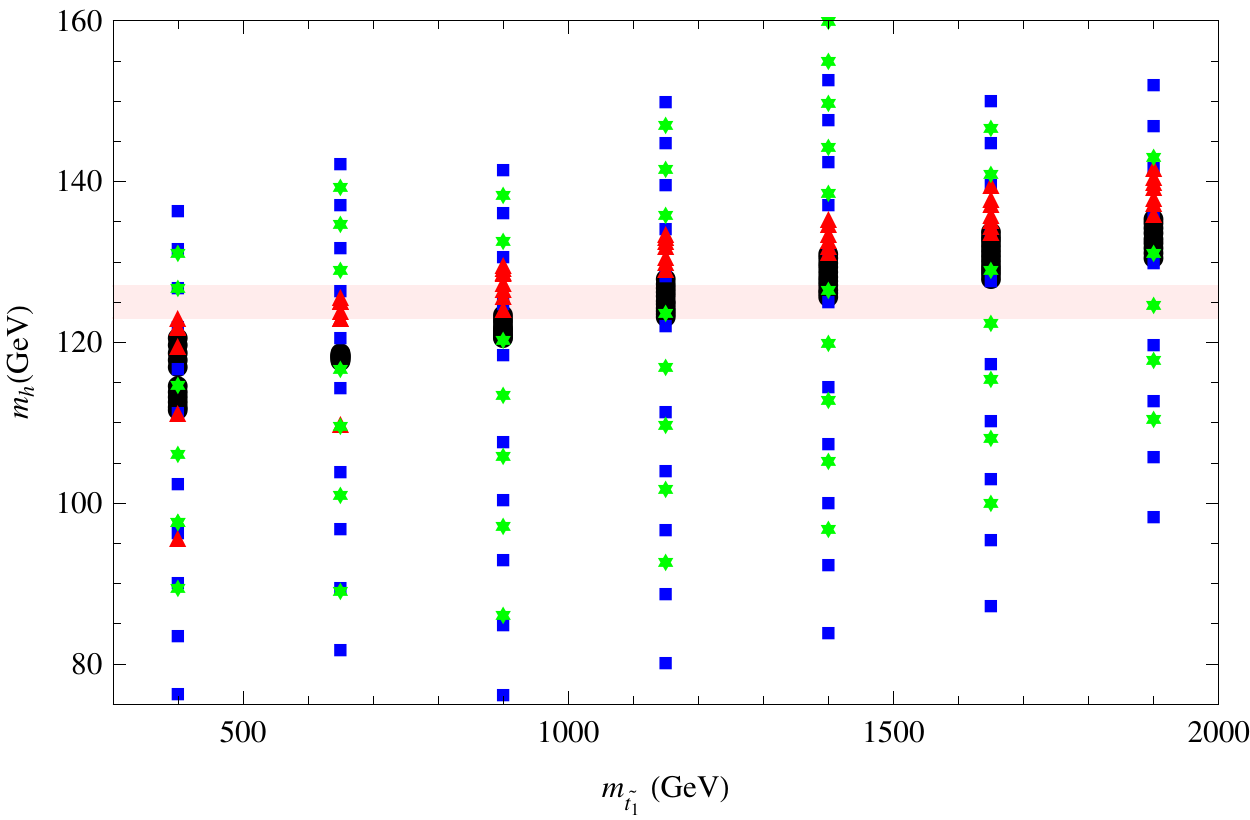} }}
\caption{The lightest Higgs mass variation with $m_{\tilde{t}_1}$ for the minimal and  maximal mixing scenarios, respectively, where 
the radiative corrections are only from the strong sector. The black and red points represent $\lambda=0.1$ whereas the green and blue points stand for $\lambda=0.9$ for $\tan\beta=5$ and $50$ respectively.}\label{xtb}
\end{figure}

Next we investigate the behaviour of the lightest Higgs mass as we vary the third generation squark masses. 
It is well known that the radiatively corrected 
Higgs mass matrix receives the most important corrections from top-stop and bottom-sbottom loops. These contributions are both proportional to the 
squark masses and the parameters in the off-diagonal mixing terms (Eqn.~\ref{stopm}). This is why we consider two different scenarios to see the effect of the 
squark mass mixing terms on the resulting Higgs mass at one loop where only the strong sector is taken into account. We consider  
first a scenario where the mixings $m_{X_t}^2$ and $m_{X_b}^2$ depend only 
on $\lambda$ term, i.e., we take $A_t=-\mu_D\cot{\beta}$ and $A_b=-\mu_D \tan{\beta}$ in Eqn.~(\ref{stopm}) as our minimal 
mixing scenario. 
Similarly the 
latter case, maximal mixing scenario, can be obtained via keeping the mixing terms as in Eqn.~(\ref{stopm}). 
In Fig.~\ref{xtb} we display the Higgs mass 
variation with the squark mass $m_{\tilde{t}_1}$ for various $\mu_D$ where the lightest Higgs mass range ($m_{h_1}=125 \pm 2$ GeV) allowed by recent CMS and ATLAS \cite{Higgsd1, Higgsd2} is shaded pink. 
In both panels we scan the parameter
space of weakly and highly coupled cases for small and large $\tan\beta$ i.e. $\lambda$=0.1, 0.9 and $\tan\beta$ =5, 50. In Fig.~\ref{xtb}(a) we consider 
the minimal mixing scenario and the black and red points represent the weakly coupled case while the green and blue points stand for the highly coupled case for 
$\tan\beta=5$ and $50$, respectively. 
In Fig.~\ref{xtb}(b) we consider the maximal mixing scenario and the color code is exactly the same as in the right panel. 
Fig.~\ref{xtb}(a) shows that when the weakly coupled case is considered for minimal mixing scenario the allowed lightest Higgs range can be reached only for 
$m_{\tilde{t}_1}\gtrsim900$ GeV.\footnote{Compare with the stop mass limit in pMSSM, which is $\gsim 3$ TeV for no-mixing scenario \cite{Djouadi}.}
This lower bound is further weakened when
the maximal mixing scenario is considered, see Fig.~\ref{xtb}(b), from where the lower bound is $\sim 700$ GeV because of the additional contributions from the MSSM parameters. In the highly coupled case
the required stop mass could be as low as 200 GeV for both the maximal and the minimal mixing scenarios.
It is seen that the triplet contribution to the third generation squark masses is crucial to decrease the required stop mass
and still have sufficient radiative corrections for the Higgs boson mass.

\begin{figure}
\centering
\mbox{\subfigure[]{\includegraphics[width=3in,height=2.5in]{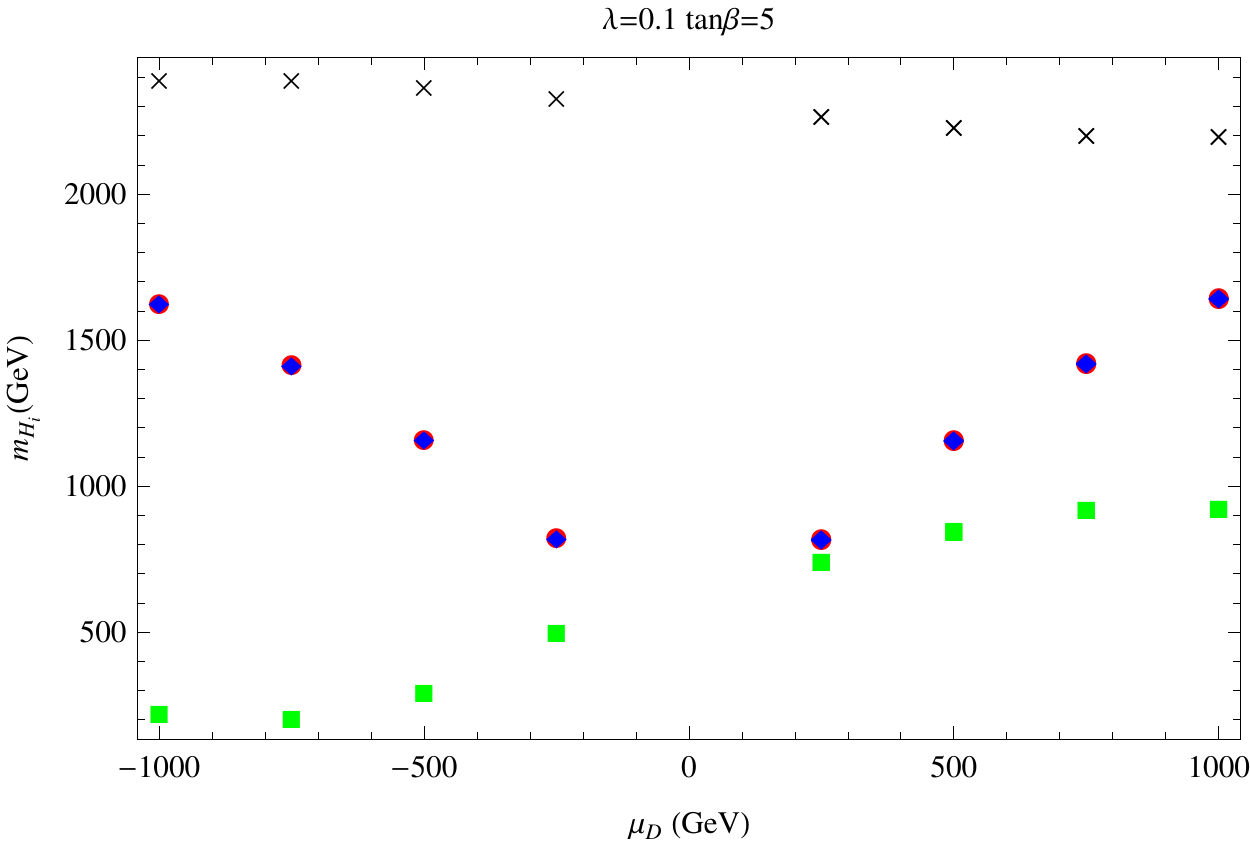}}\quad
\subfigure[]{\includegraphics[width=3in,height=2.5in]{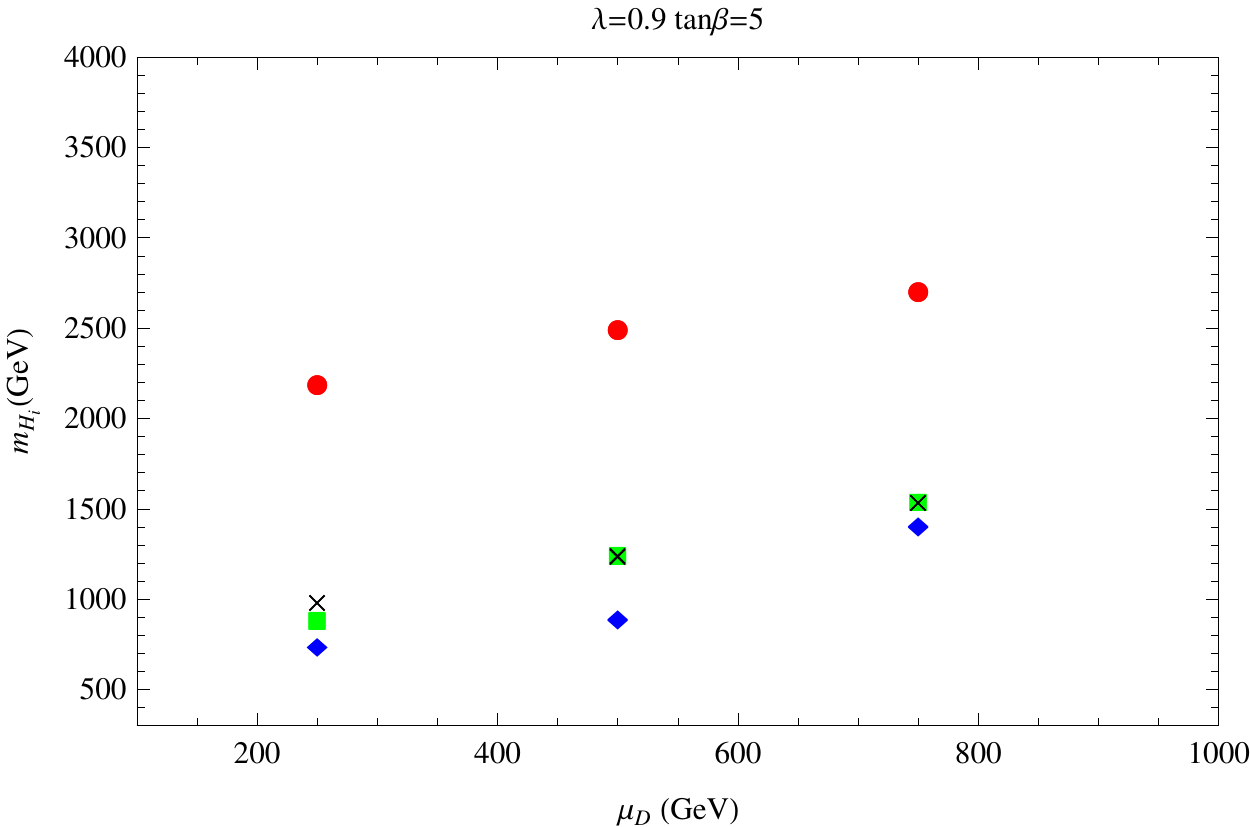} }}
\mbox{\subfigure[]{\includegraphics[width=3in,height=2.5in]{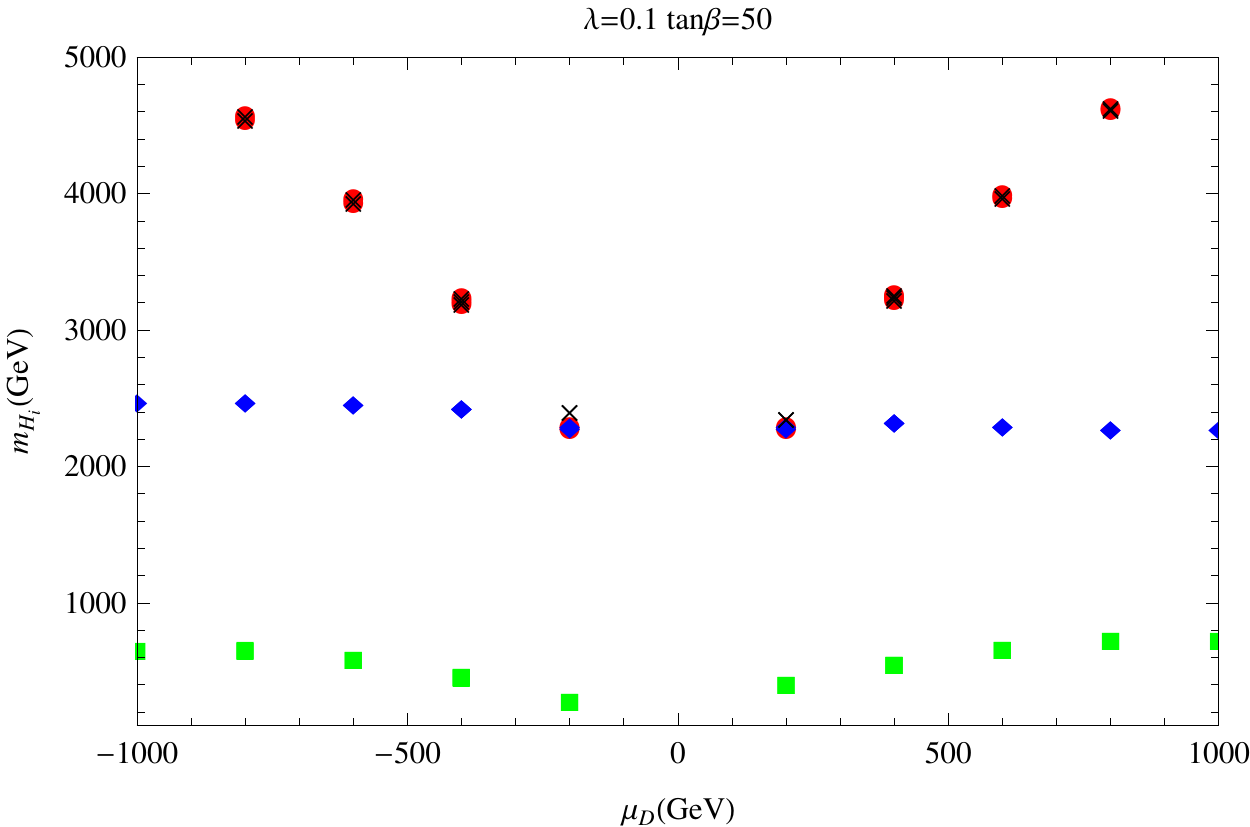}}\quad
\subfigure[]{\includegraphics[width=3in,height=2.5in]{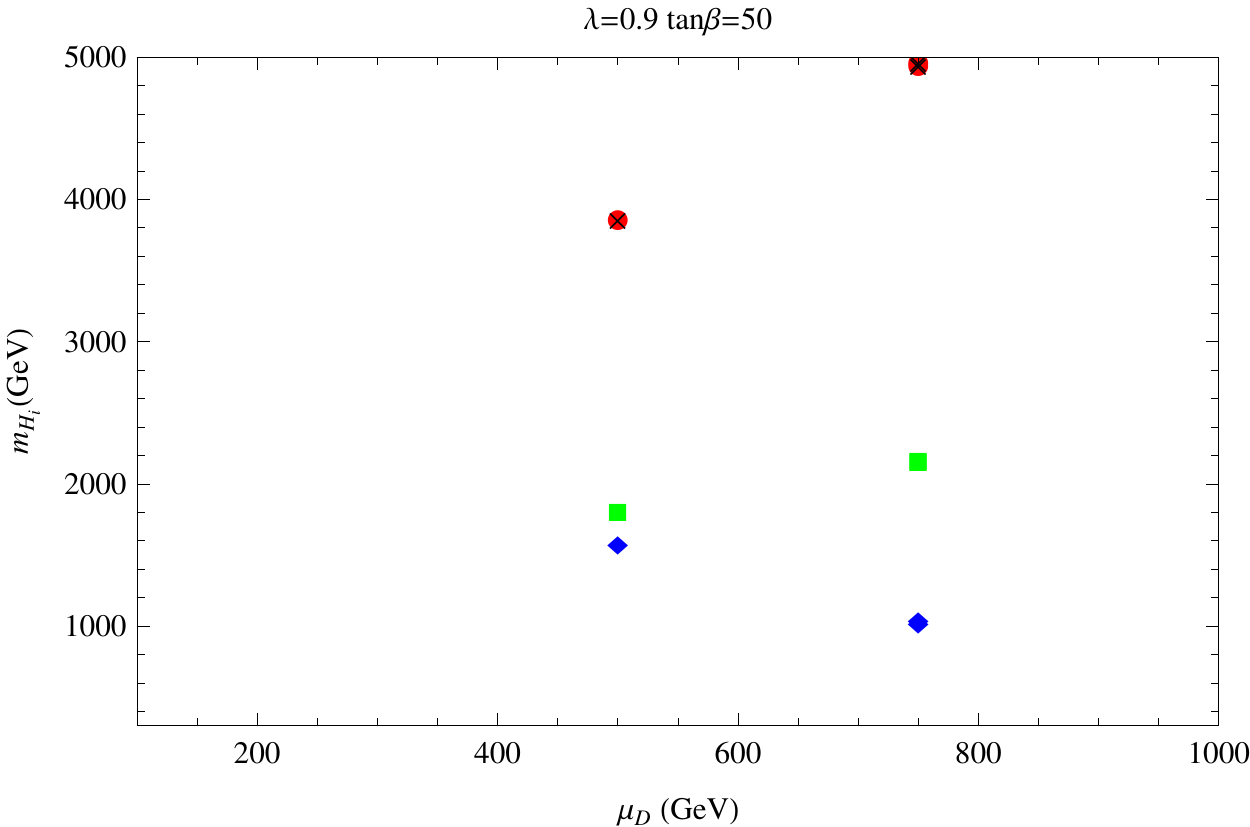} }}
\caption{Neutral Higgs spectrum versus $\mu_D$ for the scenario 2 (Sc2). Only the lightest Higgs mass is not shown in the plots. Here the masses of heavier CP-even Higgses $H_1$ and $H_2$ are displayed by the green and the red points whereas the blue and black points are for the masses of CP-odd Higgses $A_1$ and $A_2$.}
\label{higgses}
\end{figure}

In Fig.~\ref{higgses} we check the variation of the other neutral Higgs boson masses, at one-loop with $\mu_D$ for the case where each of these parameter points has $m_h\sim 125$ GeV. 
The heavy CP-even Higgses $H_{1,2}$ and CP-odd Higgses $A_{1,2}$ are 
shown for Sc2 in  Fig.~\ref{higgses}. 
The upper left plot of   Fig.~\ref{higgses} shows the variation of these Higgs boson masses for 
$\lambda=0.1$ and $\tan{\beta}=5$. We can see that for $\mu_D\lsim 0$, the second lightest neutral Higgs boson mass $m_{H_1}$ stays around 200-500 GeV. $A_1$ and $H_2$ stay degenerate for the whole parameter space whereas the heaviest pseudoscalar Higgs, $A_2$ remains decoupled at $>2$ TeV. 
Upper right plot of Fig.~\ref{higgses} corresponds to highly coupled case with triplets, i.e., $\lambda=0.9$ for $\tan{\beta}=5$. Here we 
see that $H_2$ is decoupled around $\sim 2.1-2.7$ TeV. 
The remaining Higgs masses are close to each other between 700-1600 GeV for $m_{h}\sim 125$ GeV from 
$\mu_D\sim 300-750$ GeV. 
The lower left plot of  Fig.~\ref{higgses} shows the weakly coupled case for $\tan{\beta}=50$. 
Compared to $\tan{\beta}=5$ case (upper 
left), we see that $A_1$ and $A_2$ change the behaviour, as now $A_2$ and $H_2$ are degenerate and symmetric around $\mu_D=0$. 
Unlike the $\tan{\beta}=5$ case,  both are heavier than $\sim 2.5$ TeV. 
In the lower right plot of  Fig.~\ref{higgses} the highly coupled case for $\tan{\beta}=50$ is shown. 
Compared to low $\tan{\beta}$ case (upper right), the allowed parameter space is less and  $A_2$ and $H_2$ are degenerate but decoupled with masses between 3.8-5 TeV.  

\section{Charged Higgs, chargino and $\mathcal{B}r(B_s\to X_s \gamma)$ constraint}\label{bsg}

Rare $B$-meson decay analysis provides stringent constraints on new physics beyond the Standard Model. In particular, parameter space of the MSSM like models with minimal or general flavor mixings in the sfermion sector has been investigated in great detail with the help of $B$-physics observables \cite{Bphysics}. In our analysis we calculate the inclusive radiative decay $B_s\to X_s \gamma$ and then we combine the constraints coming from Br($B_s\to X_s \gamma)$ and the constrains coming from the lightest Higgs boson mass as discussed in Section~\ref{param}.

It is known that the significant contributions to $\mathcal{B}r(B_s\to X_s \gamma)$ in the case of MSSM include the top-charged Higgs boson and stop-chargino loop contributions in addition to the SM contributions. 
The situation is similar in the Higgs triplet model, except that in general there are two more charged Higgses and one more chargino than in the MSSM. In this analysis we only consider the higgsino like charginos and also ignore the gaugino-higgsino mixing to be consistent with the analysis in Section~\ref{param}. 

The chargino sector of this model without gaugino-higgsino mixing comprise two higgsino like charginos one of which has a mass $\sim |\mu_D|$ whereas the other is around $2 |\mu_T|$. For the further analysis we ignore the $|\mu_D|\lsim 104$ GeV region because of the experimental chargino mass limit $\sim 104$ GeV \cite{charginolimit}. We take $\mu_T=1.2$ TeV so that only the light chargino contributes to the $b \to s \gamma$. We also observe that the light chargino is always dominantly doublet so its contribution to the $\mathcal{B}r(B_s\rightarrow X_s \gamma)$ will be similar with the one in the MSSM.

Compared to SM, in the case of 2HDM, the only extra contribution to $\mathcal{B}r(B_s\to X_s \gamma)$ comes from the top-charged Higgs loop, which gives a conservative lower bound to charged Higgs mass  $\gsim 230$ GeV \cite{Gambin}. In MSSM this lower bound can go down further due to the cancellation between top-charged Higgs and stop-chargino loops \cite{susybtosgamma}. 
Similarly, in our case there are possibilities for such cancellations.
The difference compared to the MSSM is that the triplet part of the charged Higgses and charginos does not couple to quarks, and thus
does not contribute to $B_s\to X_s \gamma$ decay.

\begin{figure}
\centering
\mbox{\subfigure[]{\includegraphics[width=3in,height=2.5in]{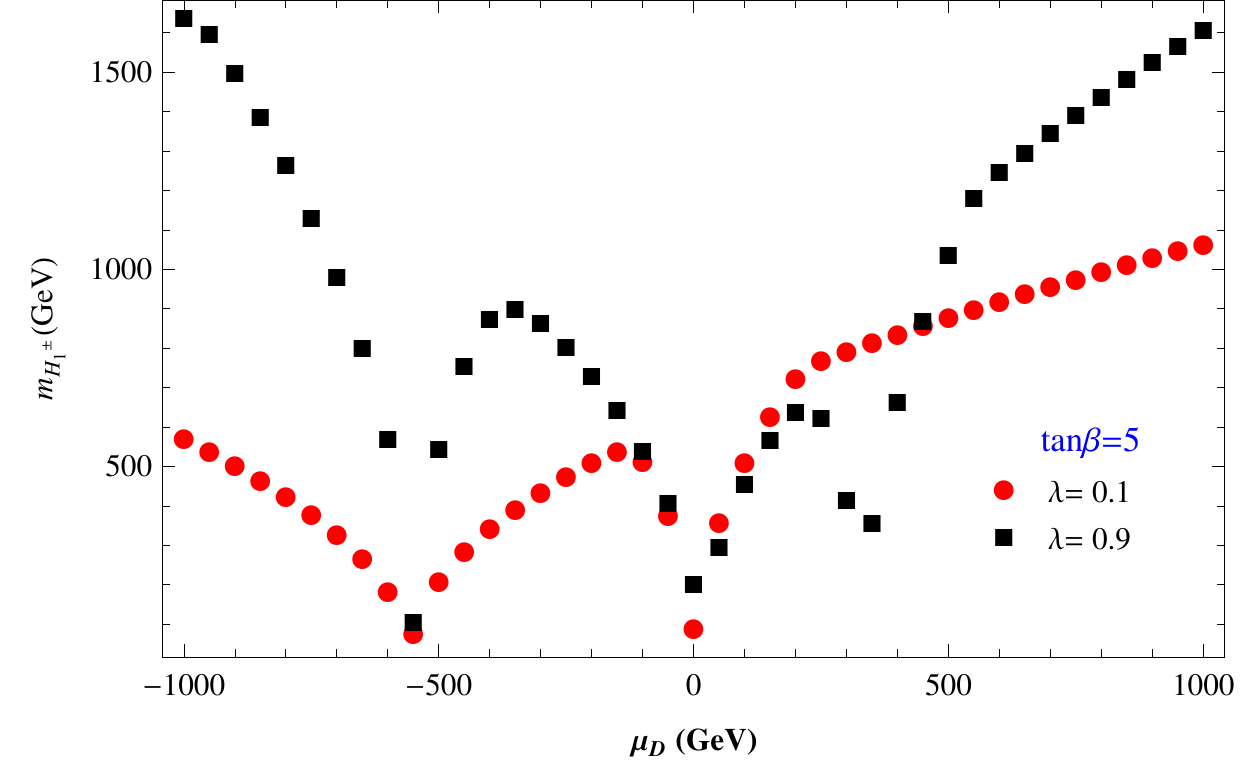}}\quad
\subfigure[]{\includegraphics[width=3in,height=2.5in]{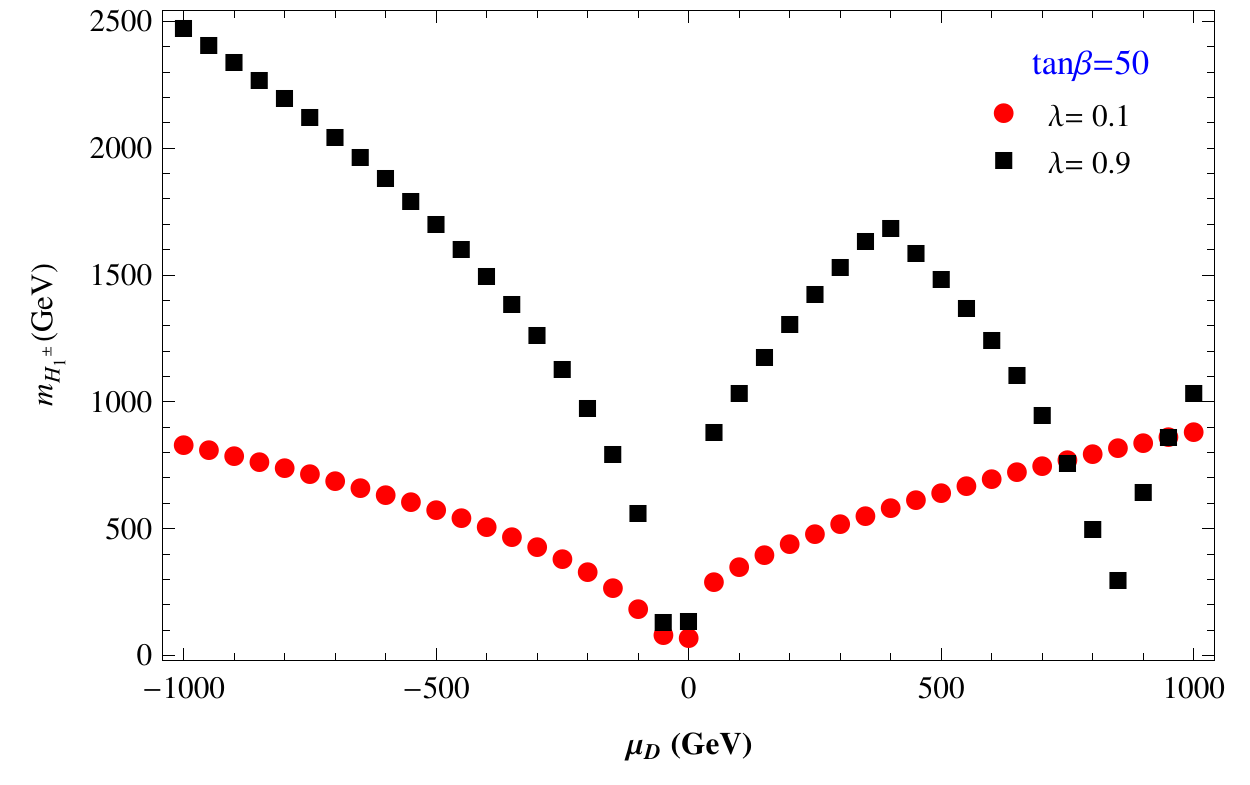} }}
\caption{Variation of lightest charged Higgs mass with  $\mu_D$ for $\lambda=0.1, 0.9$ and $\tan{\beta}=5, 50$.}\label{chhiggsmass}
\end{figure}


 Fig.~\ref{chhiggsmass} shows the lightest charged Higgs mass at tree level as a function of $\mu_D$ for the triplet mass parameter $\mu_T= 1.2$ TeV and $\tan\beta =5$ and $50$. For each $\tan\beta$, we consider the highly ($\lambda=0.9$) and weakly ($\lambda=0.1$) coupled case. 
From Fig.~\ref{chhiggsmass}, it is clear that the lightest charged Higgs is quite heavy ($\gsim 500$ GeV) for most of the parameter regions. 
The lightest charged Higgs is light ($\lsim 200$ GeV) when the doublet Higgs mixing parameter $|\mu_D|$ is small.
For $\tan{\beta}=5$, the lightest charged Higgs is light also around $\mu_D\sim -550$ GeV for this specific parameter space, because some cancellations between the doublet and the triplet terms occur in the non-diagonal terms of the charged Higgs mass matrix. In both $\tan\beta=5$ and $50$ cases when the highly coupled theory is considered for $\mu_D>0$ region, we see that the terms proportional to $\mu_D$ reduces the effects of the triplet terms in the diagonal entries of the charged Higgs mass matrix so that we can again have a light charged Higgs with mass $\leq 400$ GeV.  
For this parameter space we observed that the other two charged Higgses are mostly triplet and heavier than 1 TeV due to large $\mu_T$.
Thus, even though we have three charged Higgses in the spectrum only the lightest one affects the $B_s\to X_s \gamma$. 

The interesting situation arises when the lightest charged Higgs has a large triplet component.
The triplet nature of the charged Higgs reduces the charged Higgs contributions to $b \to s \gamma$ as discussed earlier. 
As a result $\mathcal{B}r(B_s\to X_s \gamma)$ can be very different from MSSM prediction. 
In order to estimate this effect, we study the composition of the lightest charged Higgs mass eigenstate. 

\begin{figure}
\centering
\mbox{\subfigure[]{\includegraphics[width=3.4in,height=3.0in]{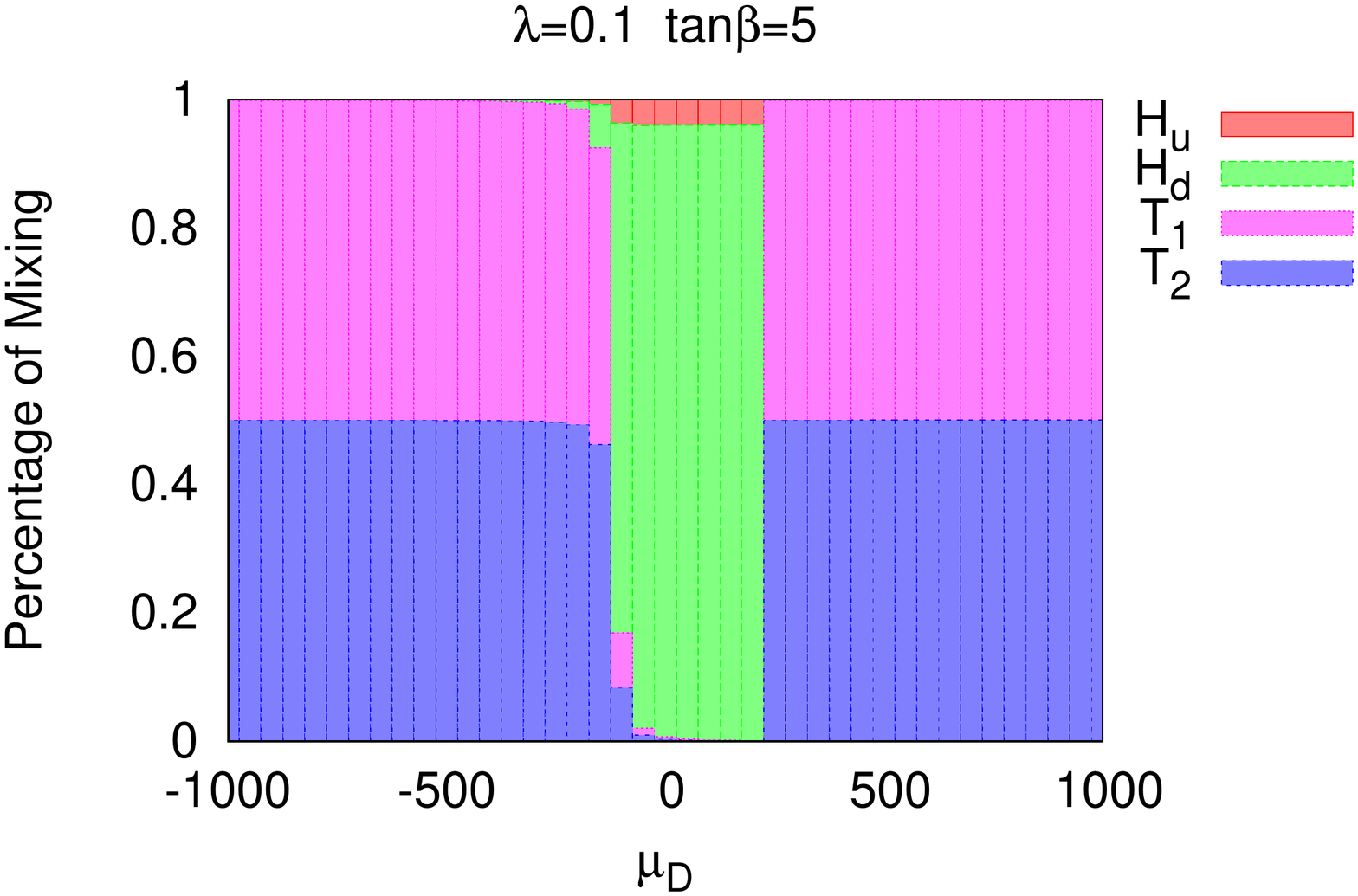}}\quad
\subfigure[]{\includegraphics[width=3.4in,height=3.0in]{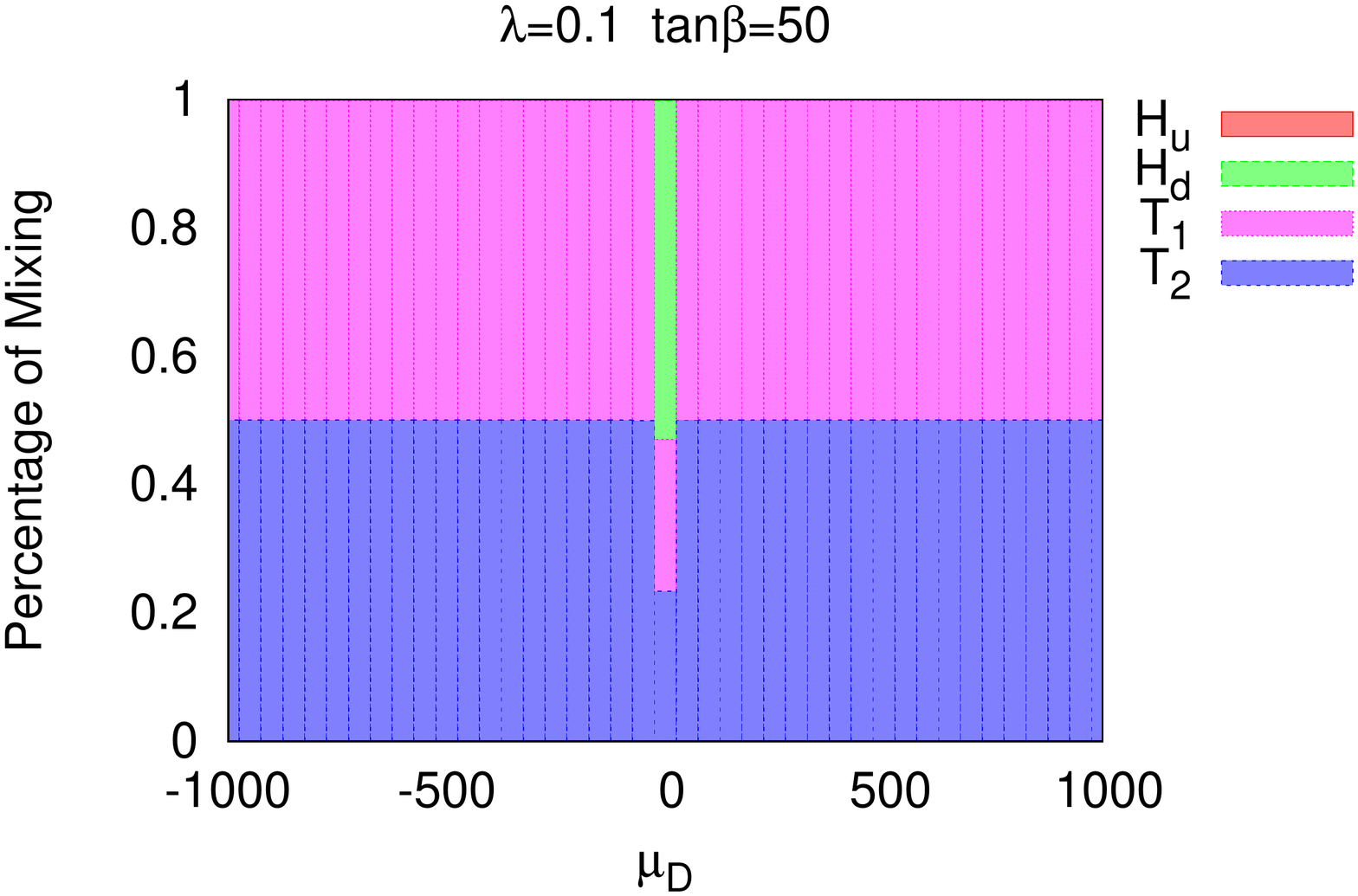} }}
\mbox{\subfigure[]{\includegraphics[width=3.4in,height=3.0in]{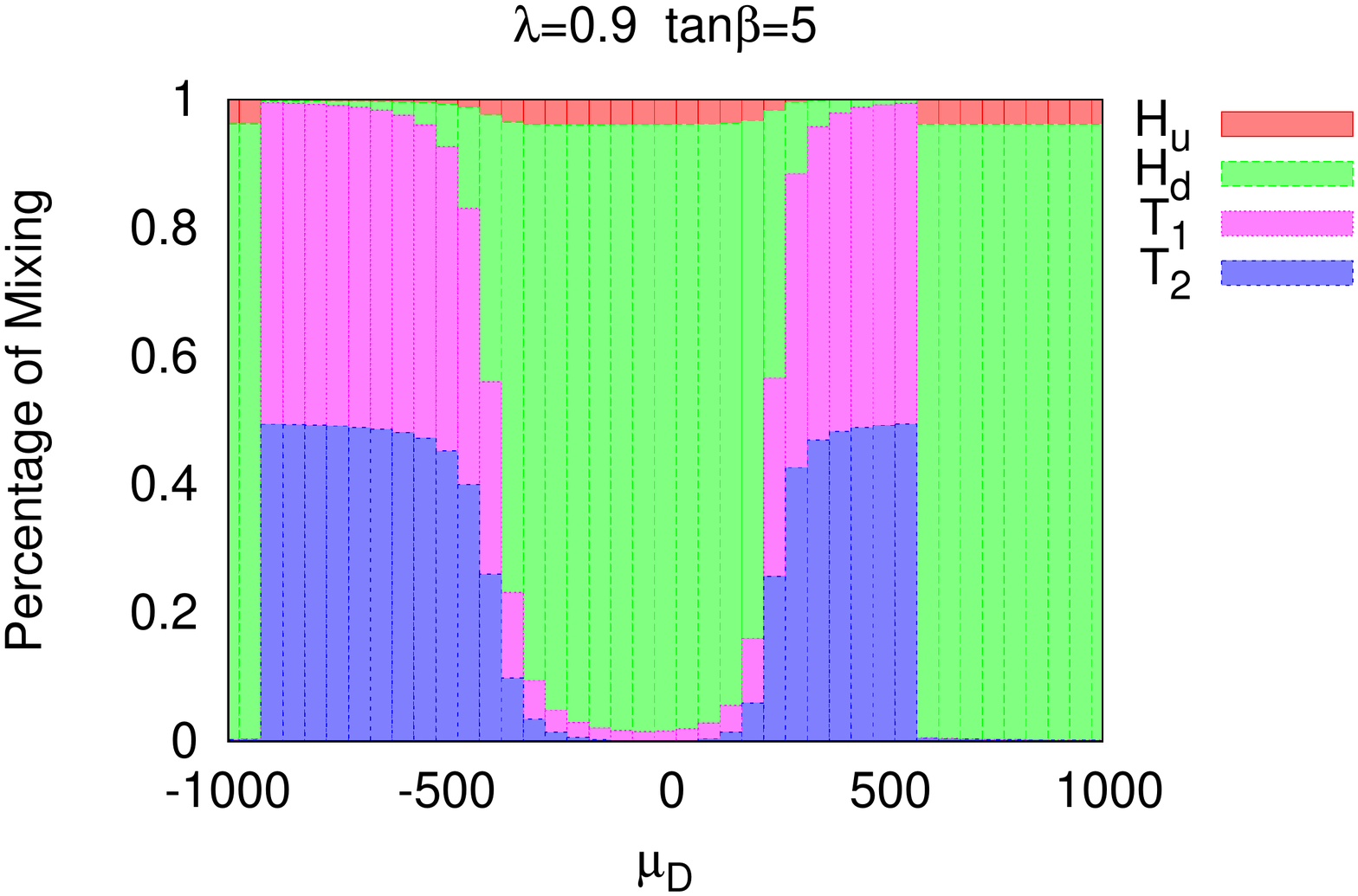}}\quad
\subfigure[]{\includegraphics[width=3.4in,height=3.0in]{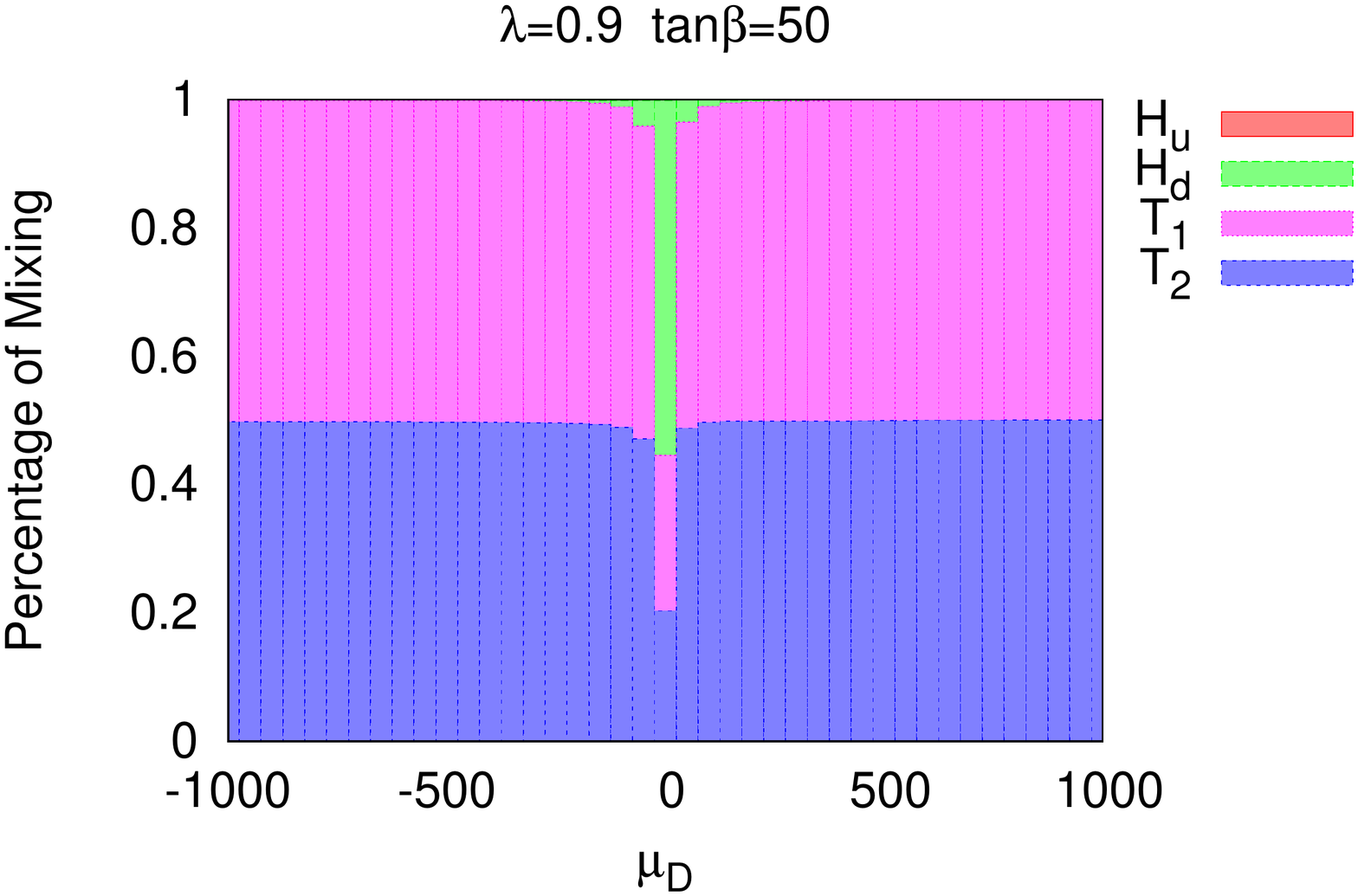} }}
\caption{The doublet-triplet in lightest charged Higgs mass eigenstate with the variation of $\mu_D$ for $\lambda=0.1, 0.9$ and $\tan{\beta}=5, 50$.}\label{chhiggsmixing} 
\end{figure}

Fig.~\ref{chhiggsmixing} presents the variation of the doublet and triplet mixing for the lightest charged Higgs in Fig.~\ref{chhiggsmass} with $\mu_D$. 
In the left top panel we display the percentage of mixing for weakly coupled theory with $\tan\beta=5$. 
The charged Higgs is generally dominantly triplet except for the $-200\leq\mu_D\leq 250$ region where it is dominantly doublet 
and contributes to $b \to s \gamma$. 
In the highly coupled theory, $\lambda=0.9$ case in the down left panel, the doublet part in the charged Higgs becomes substantial and new regions for both $\mu_D<0$ and $\mu_D>0$ become relevant for the $b \to s \gamma$. 
In the right panel we show the  probability distributions for $\tan{\beta}=50$. Unlike the low $\tan\beta$ case, the lightest charged Higgs mass eigenstate is mostly the triplet scalar field for both $\lambda=0.1$ and $0.9$. This implies the fact that the lightest charged Higgs contribution to the $\mathcal{B}r(B_s\to X_s \gamma)$ is negligible for $\tan\beta=50$.


\section{Allowed parameter space} \label{allowedbsg}
In this section we consider the constraints coming from the lightest Higgs boson mass around 125 GeV and from $\mathcal{B}r(B_s\to X_s \gamma)$ in addition to the experimental lower limit of the chargino mass. For the calculation of $\mathcal{B}r(B_s\to X_s \gamma)$, we use MicrOMEGAs version 2.4.5 \cite{micromegas} where we incorporate the percentage of mixing between doublets and triplet in the lightest charged Higgs boson of our model. 
For all the Figures in this section, the yellow band shows the allowed regions of the lightest Higgs mass $125\pm 2$ GeV; the orange band shows the LEP-excluded chargino mass region \cite{charginolimit}. 
The green points in the allowed Higgs mass band respect the experimental value of $\mathcal{B}r(B_s\to X_s \gamma)$ within $\pm 2\sigma$ \cite{btosgammaexp} where only the experimental uncertainties are taken into account. 
\begin{figure}
\centering
\mbox{\subfigure[]{\includegraphics[width=3in,height=2.5in]{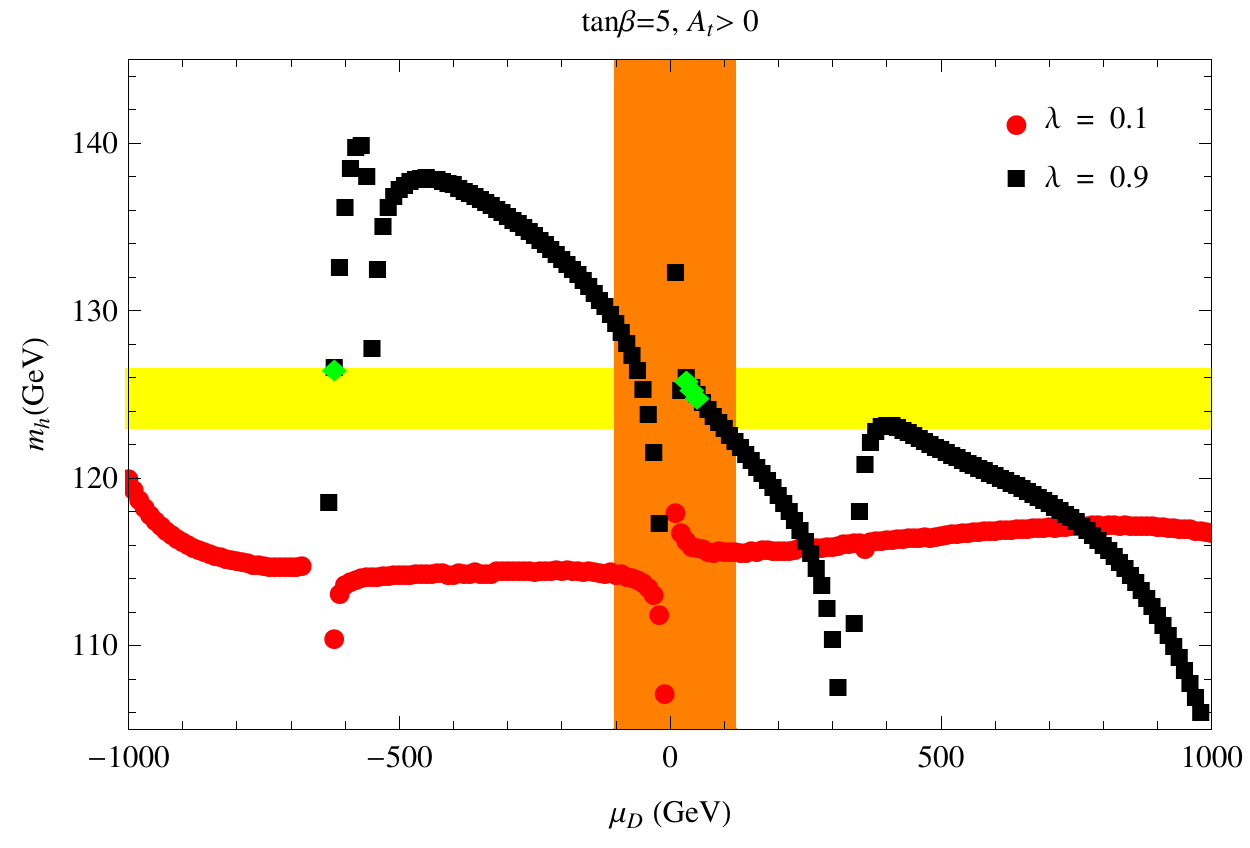}}\quad
\subfigure[]{\includegraphics[width=3in,height=2.5in]{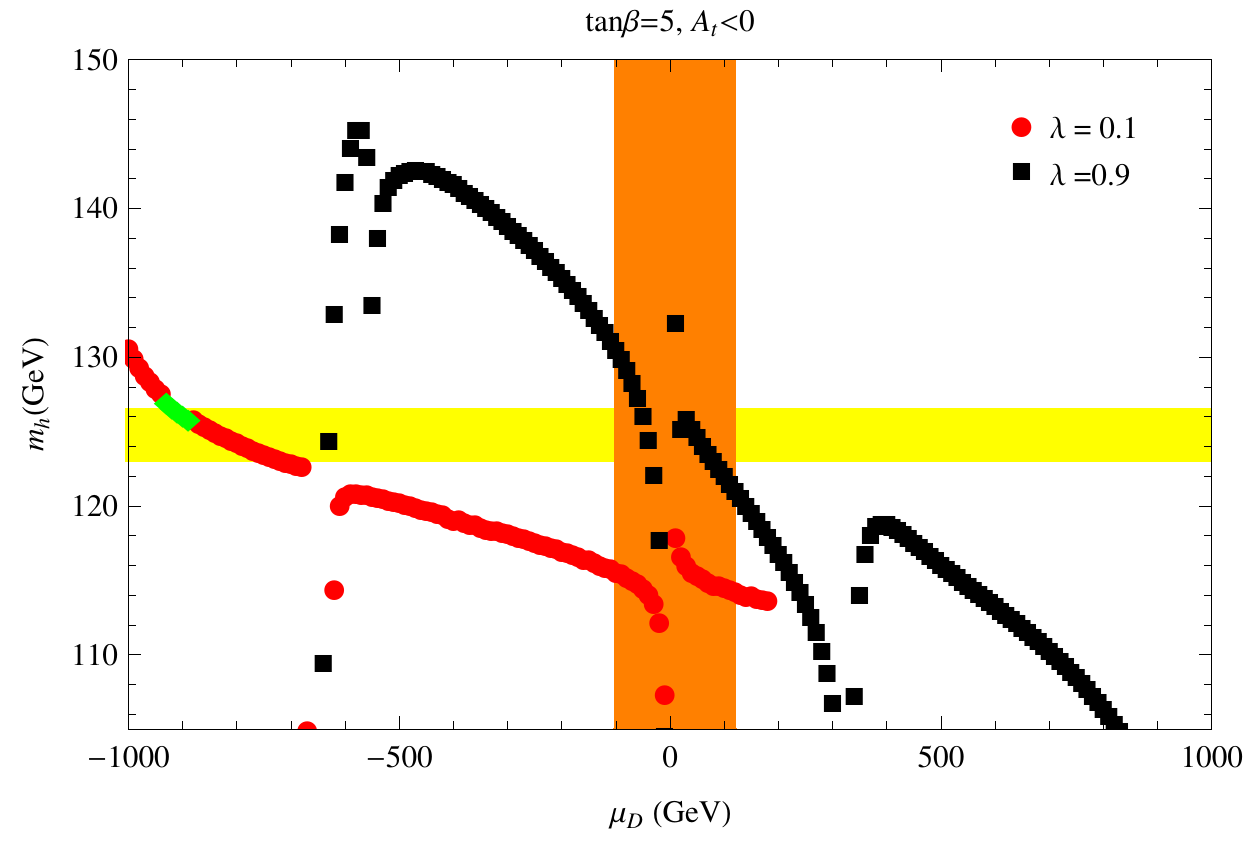} }}
\caption{Variation of $m_h$ with $\mu_D$ for  $\tan{\beta}=5$ and $m_{\tilde{t}_1}=500$ GeV when $A_t>0$ (left)  and  $A_t<0$ (right). The yellow band represents $m_h=125 \pm 2$ GeV and the orange band shows the LEP-excluded chargino mass region. The green points satisfy the allowed experimental value of $\mathcal{B}r(B_s\to X_s \gamma)$ within $\pm 2\sigma$.}\label{sc1excl}
\end{figure}

In Fig.~\ref{sc1excl} we show the variation of the lightest Higgs mass with $\mu_D$ for $\tan{\beta}=5$ and $m_{\tilde{t}_1}=500$ GeV. 
The left plot of Fig.~\ref{sc1excl} describes the case where the tri-linear coupling $A_t>0$ and the right plot of Fig.~\ref{sc1excl} shows the case of $A_t<0$. We also consider two different values of $\lambda=0.1, 0.9$ representing the weakly and highly coupled cases, respectively. 
For $A_t>0$ (left figure), for weakly coupled case, the lightest Higgs mass cannot exceed 120 GeV. 
For the $A_t<0$ case the lightest Higgs mass can reach around 130 GeV for $\lambda=0.1$ and some points around $\mu_D\sim-900$ also satisfy the experimental $\mathcal{B}r(B_s\to X_s \gamma)$ value within $\pm2\sigma$. 
In the highly coupled case for both signs of $A_t$, the lightest Higgs can be heavy and is in the allowed Higgs mass region when $|\mu_D|$ is small and when $\mu_D \sim -600$ GeV. We find that for $A_t>0$,  some points for negative $\mu_D$ values satisfy also 
$\mathcal{B}r(B_s\to X_s \gamma)$ constraint and the chargino mass limit.
For the $A_t<0$ case there are no points satisfying the  $\mathcal{B}r(B_s\to X_s \gamma)$ constraint within $\pm2 \sigma$. However we observe that the  $\mathcal{B}r(B_s\to X_s \gamma)$ values for these points are very close $\pm2\sigma$ region. If we assume that the theoretical uncertainties on $\mathcal{B}r(B_s\to X_s \gamma)$ for this model are similar with the MSSM ones \cite{btosgammauncertainities} then the inclusion of theoretical uncertainties of the order of one sigma would bring several points into the allowed region.

\begin{figure}
\centering
\mbox{\subfigure[]{\includegraphics[width=3in,height=2.5in]{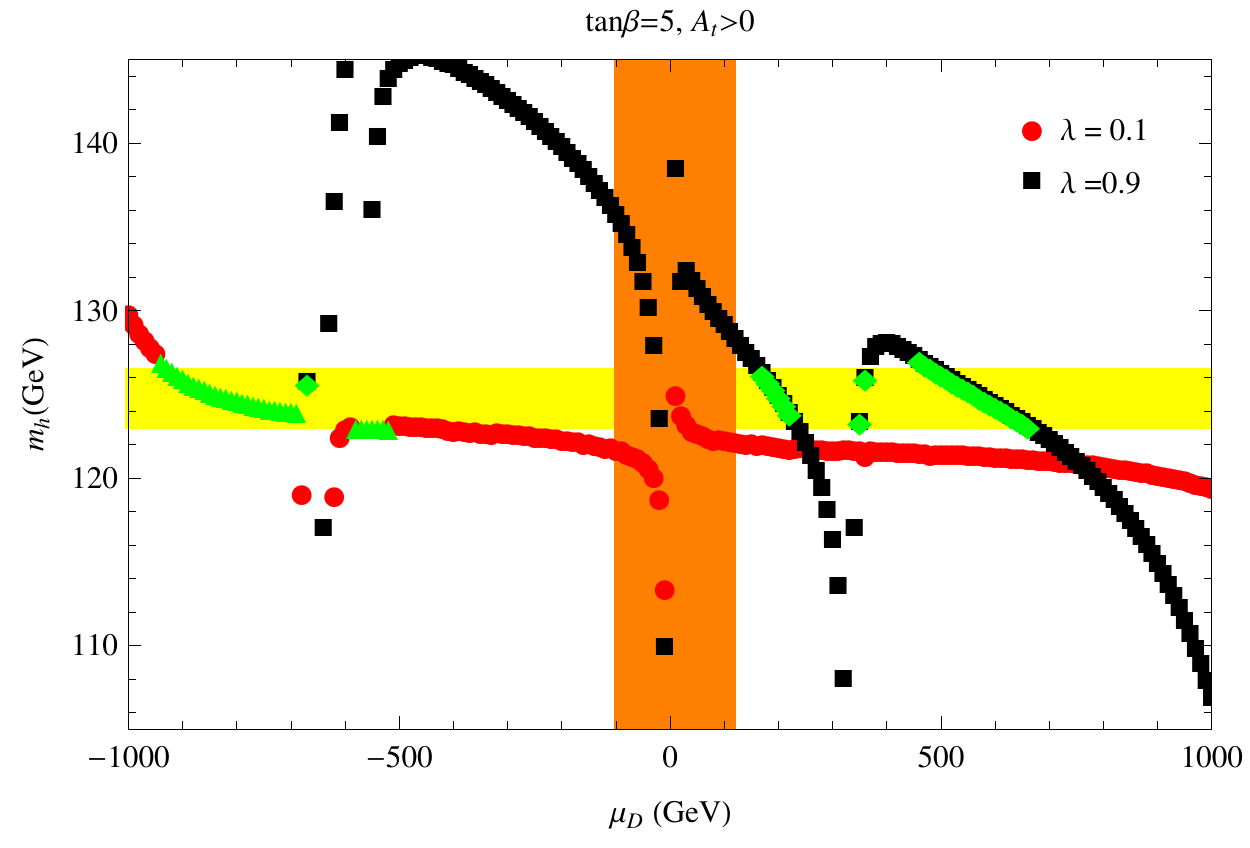}}\quad
\subfigure[]{\includegraphics[width=3in,height=2.5in]{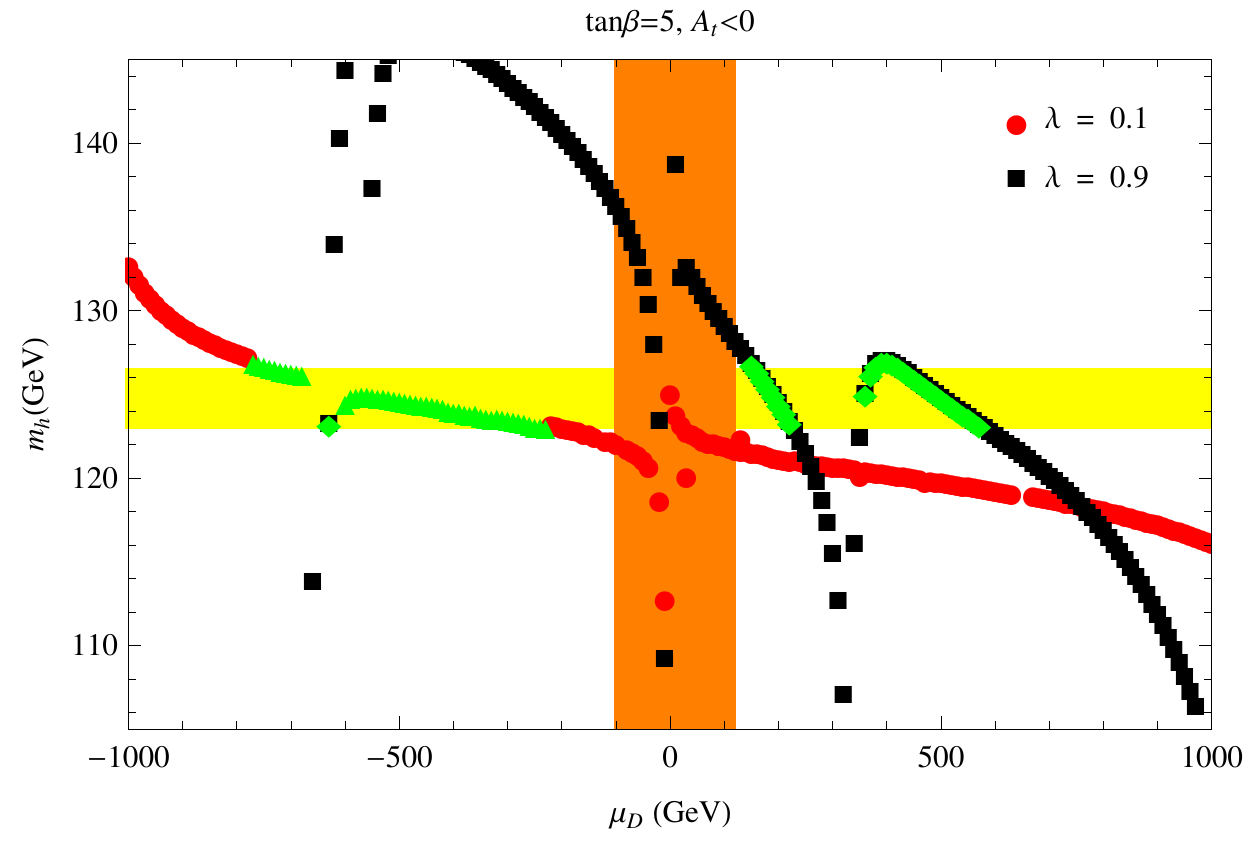} }}
\caption{Otherwise same as Fig.~11 but $m_{\tilde{t}_1}=1$ TeV.}
\label{sc1excl1tev}
\end{figure}

Fig.~\ref{sc1excl1tev} shows the case for $m_{\tilde{t}_1}=1$ TeV for both $A_t>0$ and $A_t<0$. For heavier ${\tilde{t}_1}$ more points are allowed for highly coupled case compared to the $m_{\tilde{t}_1}=500$ GeV case. For $\lambda =0.9$ the variations of Higgs mass with $\mu_D$ do not differ much for  $A_t>0$ and for $A_t<0$. The lightest Higgs mass is largely within the allowed band for $\mu_D \sim 100-700$ GeV. In the weakly coupled case for both $A_t>0$ and $A_t<0$ there are a number of points with $m_h\sim 125$ GeV for $\mu_D <0$, compared to the earlier case of $m_{\tilde{t}_1}\sim 500$ GeV, where only  $A_t<0$ has some allowed points. The difference between $A_t>0$ and $A_t<0$ cases is mainly that for $A_t>0$ the allowed $\mu_D$ values are relatively smaller than for  the $A_t<0$ case.

Next we consider $\tan{\beta}=50$ case, where  the lightest charged Higgs is mostly triplet (see Fig.~\ref{chhiggsmixing}) and thus does not contribute significantly to the $b\to s \gamma $ decay. For large $\tan{\beta}$ the chargino contribution increases, specially for the relatively light stop, $m_{\tilde{t}_1}\sim 500$ GeV. We have investigated the case of $m_{\tilde{t}_1}=500$ GeV for $\tan{\beta}=50$ and found that the chargino contribution is comparable to the SM contribution \footnote{
 However, the numerical calculation cannot be trusted in this case. One needs $\alpha_s^2$ NLO corrections, which are not currently available in the codes we have used. The problem has been verified with both micrOMEGAs and CPsuperH authors \cite{JLMN}.}. 

\begin{figure}
\centering
\mbox{\subfigure[]{\includegraphics[width=3in,height=2.5in]{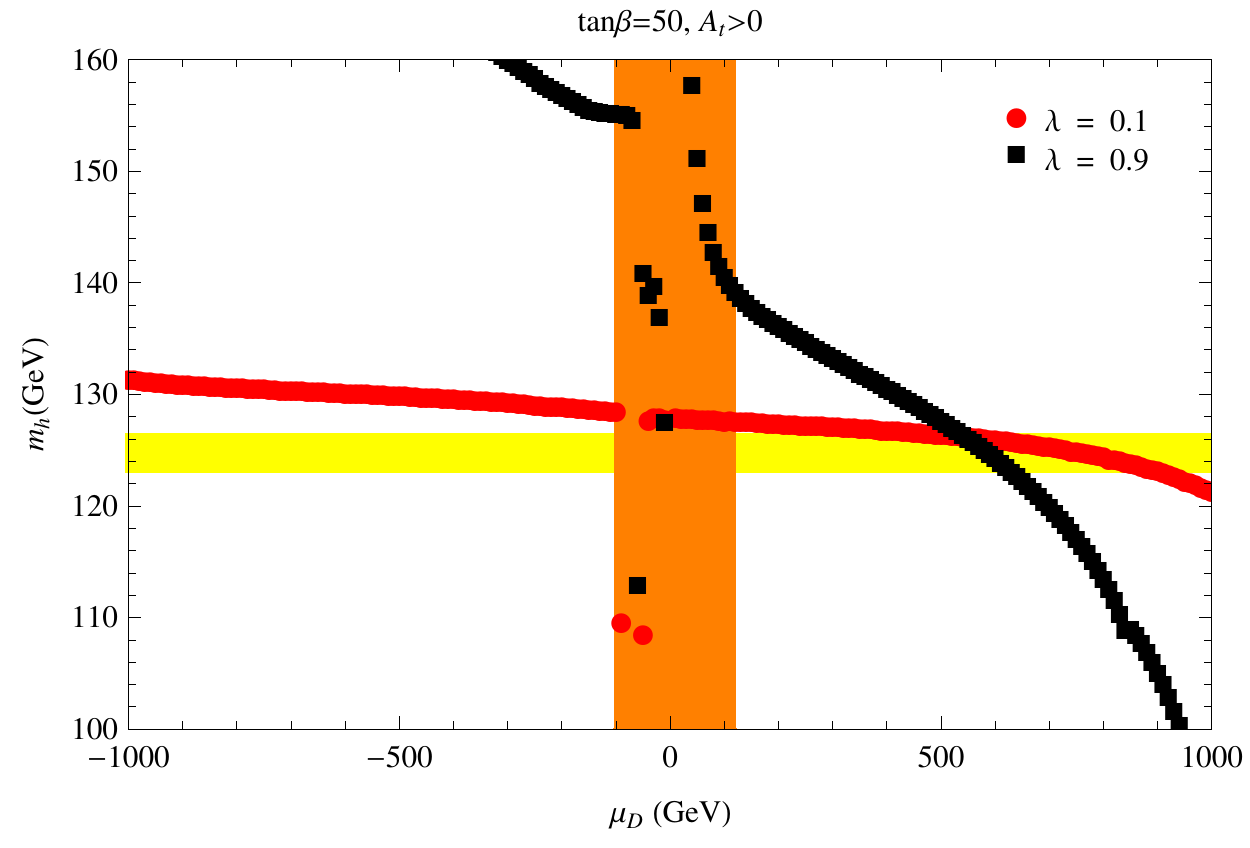}}\quad
\subfigure[]{\includegraphics[width=3in,height=2.5in]{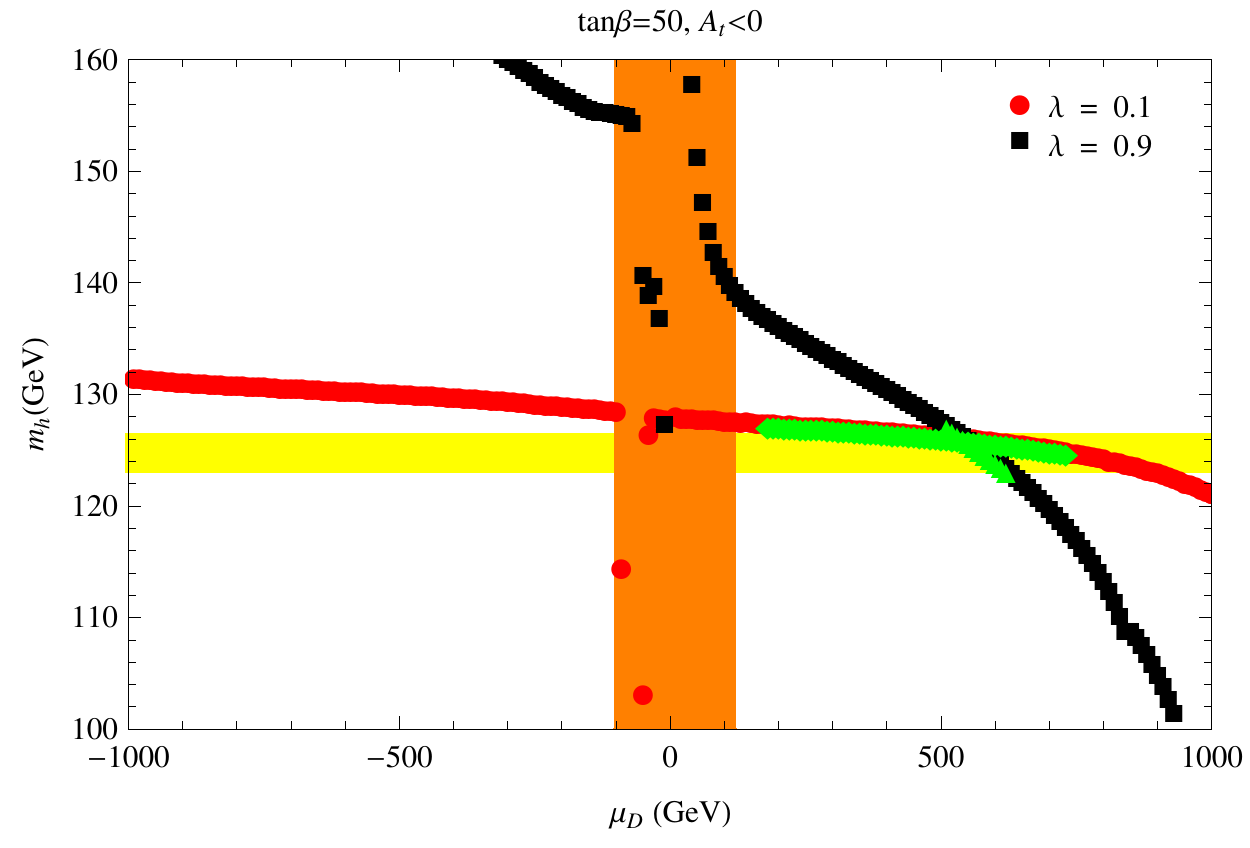} }}
\caption{Otherwise same as Fig.~11, but $\tan{\beta}=50$ and $m_{\tilde{t}_1}=1$ TeV.}
\label{sc2excl1tev}
\end{figure}

In the case of $m_{\tilde{t}_1}=1$ TeV and $\tan{\beta}=50$ the situation is very different, since the stop-chargino contributions reduce due to heavy stop. 
Fig.~\ref{sc2excl1tev} shows that the Higgs mass variations in the weakly coupled case for both the $A_t>0$ and $A_t<0$ cases look similar and many points satisfy the $\mathcal{B}r(B_s\to X_s \gamma)$ constraint within $\pm2\sigma$ for $A_t<0$. 
In the highly coupled case as well points satisfying the $\mathcal{B}r(B_s\to X_s \gamma)$ constraint are found for $A_t<0$.
The $\mathcal{B}r(B_s\to X_s \gamma)$ constraint prefers the negative sign  of $A_t$ over the positive sign for both the weakly and highly coupled cases. 
For $A_t>0$ many allowed points could be found if theoretical uncertainties were taken into account within two sigma.

\section{Conclusions}\label{conclu}
We have studied the triplet extended supersymmetric model in the context of 125 GeV Higgs. In TESSM the lightest Higgs with $m_h \sim 125$ GeV, does not strongly constrain the 
third generation squark masses, the trilinear couplings $A_q$, the supersymmetric bi-linear Higgs mixing term $\mu_D$ or $\tan{\beta}$ due to the addition of two triplet parameters:  triplet Higgs coupling $\lambda$, and triplet mass $\mu_T$. We have found that it is possible to achieve 
 $m_h\sim 125$ GeV even in the minimal mixing scenario  for large coupling $\lambda$ with the third generation quark masses $m_{\tilde{t}}\geq 400$ GeV.
It was also shown that the sign of $\mu_D$ depends strongly on the choice of the parameter space in general. When the third generation squark masses are light $\sim 500$ GeV, $m_h\sim 125$ GeV can be achieved as long as $\mu_D<0$. However both signs of $\mu_D$ are opened for the third generation masses $\sim 1$ TeV depending on the value of $\tan\beta$.

We have analyzed the parameter space in detail by combining the constraints coming from the allowed $\sim 125$ GeV Higgs with the constrains from $\mathcal{B}r(B_s\to X_s \gamma)$. We emphasized that it is possible to obtain many allowed points for both signs of $\mu_D$ and various $\tan\beta$ even when only the experimental uncertainties on the $\mathcal{B}r(B_s\to X_s \gamma)$ are taken into account. 

To test the characteristics of the lightest Higgs, one needs to explore all its decay modes. 
As discussed in the introduction, the mass peak in the Higgs decays can be seen in the decays to $H\to Z Z$ and $\gamma \gamma$.
The other experimentally observed decay modes are Higgs to $WW$, $b\bar{b}$, and $\tau\bar{\tau}$.
All the branching ratios must agree with a proposed model, but the study of compatibility needs still more statistics.
There is also a possibility of the lightest Higgs decaying into lightest neutralino pairs, contributing to invisible decay width of the lightest Higgs boson.

The other neutral Higgses in TESSM remain decoupled for $m_h\sim 125$ GeV. 
In TESSM, there are two CP-odd physical neutral Higgses, phenomenology of which is quite interesting. 
One has to search for them at the LHC with 14 TeV as we find that they are quite heavy. 
There are also two charged Higgses which is very different from MSSM. 
In particular, the doublet and triplet mixture in the charged Higgs spectrum leads to some interesting collider signatures at the LHC \cite{PBKHAS}. 

\vspace*{1cm}

{\bf Acknowledgments:} 
The authors gratefully acknowledge support from the Academy of Finland (Project No. 137960). The work of AS is also supported by the GRASPANP and the Finnish Cultural Foundation. PB thanks Jae Sik Lee, Matthias Neubert, Lasse Leinonen, Timo Ruppell,  Harri Waltari and Sourav Bhattacharya for various important discussions in the course of the project. 

\section{Appendix}\label{appendix}
The scalar Higgs potential of the TESSM is 
\begin{eqnarray}
V=V_F+V_D+V_S
\end{eqnarray}
where the $V_F$ can be obtained from the superpotential in Eqn.~(\ref{superpot}) as 
\begin{eqnarray}
V_F & = &  \bigg| \mu_D (v_u+h_{ur}+ih_{ui}) + \lambda \bigg( H_u^+ \xi^-_1 - {1 \over \sqrt{2}} (v_u+h_{ur}+ih_{ui}) \xi^0 \bigg) \bigg|^2 \cr
&+&  \bigg| \mu_D (v_d+h_{dr}+ih_{di}) + \lambda \bigg( H_d^- \xi^+_2 - {1 \over \sqrt{2}} (v_d+h_{dr}+ih_{di}) \xi^0 \bigg) \bigg|^2  \cr
&+& \bigg| \mu_D H_u^+ + \lambda \bigg( {1 \over \sqrt{2}} H_u^+ \xi^0 - (v_u+h_{ur}+ih_{ui}) \xi_2^+ \bigg) \bigg|^2 \cr
&+& \bigg| \mu_D H_d^- + \lambda \bigg( {1 \over \sqrt{2}} H_1^- \xi^0 - (v_d+h_{dr}+ih_{di}) \xi^-_1 \bigg) \bigg|^2  \cr
& + & \bigg|2 \mu_T \xi^0 - {\lambda \over \sqrt{2}} \bigg( (v_d+h_{dr}+ih_{di})(v_u+h_{ur}+ih_{ui}) + H_d^- H^+_u \bigg) \bigg|^2  \cr
&+& \bigg| \lambda  (v_d+h_{dr}+ih_{di}) H^+_u - 2 \mu_T \xi_2^+ \bigg|^2 + \bigg| \lambda H_d^- (v_u+h_{ur}+ih_{ui}) - 2 \mu_T \xi_1^- \bigg|^2 
\end{eqnarray}
and the D term contributions are given as  
\begin{eqnarray}
V_{D} & = & {g^2 \over 8} \bigg[ (v_d+h_{dr})^2 +h_{di}^2 - |H_d^-|^2 + |H^+_u|^2 - (v_u+h_{ur})^2 - h_{ui}^2 + 2 |\xi^+_2|^2 - 2|\xi_1^-|^2 \bigg]^2  \cr
& + & {g'^2 \over 8} \bigg[ (v_d+h_{dr})^2 +h_{di}^2 + |H_d^-|^2 - |H^+_u|^2 - (v_u+h_{ur})^2 - h_{ui}^2  \bigg]^2    \cr
& + & {g^2 \over 8} \bigg[ (v_d+h_{dr}-ih_{di}) H_d^- + H^{+*}_u (v_u+h_{ur}+ih_{ui}) + \sqrt{2} ( \xi^+_2 + \xi_1^- ) \xi^{0*} + {\rm H.c.} \bigg]^2 \cr
& -& {g^2 \over 8} \bigg[ H_d^{-*} (v_d+h_{dr}+ih_{di}) + (v_u+h_{ur}-ih_{ui}) H_u^+ + \sqrt{2} ( \xi^+_2 - \xi_1^- ) \xi^{0*} - {\rm H.c.} \bigg]^2.
\end{eqnarray}
Here the conventions used for $H_d$ and $H_u$ are
\begin{eqnarray}
H_d = \left(\begin{array}{c}
v_d+\frac{1}{\sqrt2}(h_{dr}+ih_{di})\\
H_d^- \end{array}\right),~~~~~~~~H_u= \left(
\begin{array}{c}
H_u^+ \\
v_u+\frac{1}{\sqrt2}(h_{ur}+ih_{ui})
\end{array}
\right)
\end{eqnarray}


\begin{thebibliography}{99}

\bibitem{Higgsd1}
  S.~Chatrchyan {\it et al.}  [CMS Collaboration],
  Phys.\ Lett.\ B {\bf 716} (2012) 30
  [arXiv:1207.7235 [hep-ex]].

\bibitem{Higgsd2}
  G.~Aad {\it et al.}  [ATLAS Collaboration],
  Phys.\ Lett.\ B {\bf 716} (2012) 1
  [arXiv:1207.7214 [hep-ex]].

\bibitem{ZZexcess}
[ATLAS Collaboration],
ATLAS-CONF-2013-013;
[CMS Collaboration], CMS-PAS-HIG-13-002.

%
\bibitem{newmasses}
 [ATLAS Collaboration],
  ATLAS-CONF-2013-014;
[CMS Collaboration], CMS-PAS-HIG-13-005.


\bibitem{Yang}
  C.~-N.~Yang,
  Phys.\ Rev.\  {\bf 77} (1950) 242.

\bibitem{cMSSM}
  A.~H.~Chamseddine, R.~L.~Arnowitt and P.~Nath,
  Phys.\ Rev.\ Lett.\  {\bf 49} (1982) 970;
  G.~L.~Kane, C.~F.~Kolda, L.~Roszkowski and J.~D.~Wells,
  Phys.\ Rev.\ D {\bf 49} (1994) 6173
  [hep-ph/9312272].

\bibitem{cmssmstd}
  H.~Baer, V.~Barger and A.~Mustafayev,
  JHEP {\bf 1205} (2012) 091
  [arXiv:1202.4038 [hep-ph]];

  J.~Ellis and K.~A.~Olive,
  Eur.\ Phys.\ J.\ C {\bf 72} (2012) 2005
  [arXiv:1202.3262 [hep-ph]];
 P.~Nath,
  Int.\ J.\ Mod.\ Phys.\ A {\bf 27}, 1230029 (2012)
  [arXiv:1210.0520 [hep-ph]].

\bibitem{mssmsd} 
  M.~Carena, S.~Gori, N.~R.~Shah and C.~E.~M.~Wagner,
  JHEP {\bf 1203}, 014 (2012)
  [arXiv:1112.3336 [hep-ph]].

\bibitem{naturalsusy}
  L.~J.~Hall, D.~Pinner and J.~T.~Ruderman,
  JHEP {\bf 1204} (2012) 131
  [arXiv:1112.2703 [hep-ph]].


\bibitem{cms4l}
CMS-PAS-HIG-13-002
\bibitem{atlas4l}
ATLAS-CONF-2013-013

\bibitem{cms2g}
CMS-PAS-HIG-13-001
\bibitem{atlas2g}
ATLAS-CONF-2013-012
\bibitem{baryon}
  M.~B.~Gavela, P.~Hernandez, J.~Orloff and O.~Pene,
  Mod.\ Phys.\ Lett.\ A {\bf 9} (1994) 795
  [hep-ph/9312215];
  M.~B.~Gavela, P.~Hernandez, J.~Orloff, O.~Pene and C.~Quimbay,
  Nucl.\ Phys.\ B {\bf 430} (1994) 382
  [hep-ph/9406289].

\bibitem{EDMs}
  J.~R.~Ellis, S.~Ferrara and D.~V.~Nanopoulos,
  Phys.\ Lett.\ B {\bf 114} (1982) 231;
  J.~Polchinski and M.~B.~Wise,
  Phys.\ Lett.\ B {\bf 125} (1983) 393.

\bibitem{SCPVsinglet}
 A.~Pomarol,
  Phys.\ Rev.\ D {\bf 47} (1993) 273
  [hep-ph/9208205]; 
  M.~Masip and A.~Rasin,
  Phys.\ Rev.\ D {\bf 58} (1998) 035007
  [hep-ph/9803271],

\bibitem{SCPVtriplet}
  G.~Barenboim and J.~Bernabeu,
  Z.\ Phys.\ C {\bf 73} (1997) 321
  [hep-ph/9603379];
  G.~Barenboim, M.~Gorbahn, U.~Nierste and M.~Raidal,
  Phys.\ Rev.\ D {\bf 65} (2002) 095003
  [hep-ph/0107121].
   S.~W.~Ham and S.~K.~OH,
  arXiv:0812.1419 [hep-ph].

\bibitem{SCPVMSSM}
  A.~Pomarol,
  Phys.\ Lett.\ B {\bf 287} (1992) 331
  [hep-ph/9205247].



\bibitem{tripletlit}
  H.~Georgi and M.~Machacek,
  Nucl.\ Phys.\ B {\bf 262} (1985) 463.
  M.~S.~Chanowitz and M.~Golden,
  Phys.\ Lett.\ B {\bf 165} (1985) 105.

  J.~F.~Gunion, R.~Vega and J.~Wudka,
  Phys.\ Rev.\ D {\bf 42} (1990) 1673.

  J.~F.~Gunion, R.~Vega and J.~Wudka,
  Phys.\ Rev.\ D {\bf 43} (1991) 2322.

\bibitem{PDG}
  J.~Beringer {\it et al.}  [Particle Data Group Collaboration],
  Phys.\ Rev.\ D {\bf 86} (2012) 010001.

\bibitem{triplet} 

 J. R. Espinosa and M. Quiros, 
 Nucl. Phys. B {\bf 384} (1992) 113;
 J. R. Espinosa and M. Quiros, 
 Phys. Lett. B {\bf 279} (1992) 92.


\bibitem{rho} 
 S. D. Chiara and K. Hsieh, 
 Phy. Rev. D {\bf 78} (2008) 055016.

 \bibitem{chargedH}
  J.~L.~Diaz-Cruz, J.~Hernandez-Sanchez, S.~Moretti and A.~Rosado,
  Phys.\ Rev.\ D {\bf 77} (2008) 035007
  [arXiv:0710.4169 [hep-ph]].

\bibitem{Delgado:2012sm}
  A.~Delgado, G.~Nardini and M.~Quiros,
  Phys.\ Rev.\ D {\bf 86} (2012) 115010
  [arXiv:1207.6596 [hep-ph]].

\bibitem{Delgado:2013zfa}
  A.~Delgado, G.~Nardini and M.~Quiros,
  arXiv:1303.0800 [hep-ph].

\bibitem{coleman-weinberg}
 S.~R.~Coleman and E.~J.~Weinberg,
  Phys.\ Rev.\ D {\bf 7} (1973) 1888.

\bibitem{mtop}
  T.~Aaltonen {\it et al.}  [CDF and D0 Collaborations],
  Phys.\ Rev.\ D {\bf 86} (2012) 092003
  [arXiv:1207.1069 [hep-ex]].

\bibitem{PBKHAS} 
Work in progress.

\bibitem{Djouadi}
  A.~Arbey, M.~Battaglia, A.~Djouadi, F.~Mahmoudi and J.~Quevillon,
  Phys.\ Lett.\ B {\bf 708} (2012) 162
  [arXiv:1112.3028 [hep-ph]].

\bibitem{Bphysics}
  A.~J.~Buras, P.~H.~Chankowski, J.~Rosiek and L.~Slawianowska,
  Nucl.\ Phys.\ B {\bf 659} (2003) 3
  [hep-ph/0210145]; 
  J.~Foster, K.~-i.~Okumura and L.~Roszkowski,
  JHEP {\bf 0508} (2005) 094
  [hep-ph/0506146]; 
  M.~S.~Carena, A.~Menon, R.~Noriega-Papaqui, A.~Szynkman and C.~E.~M.~Wagner,
  Phys.\ Rev.\ D {\bf 74} (2006) 015009
  [hep-ph/0603106]; 
  A.~Freitas, E.~Gasser and U.~Haisch,
  Phys.\ Rev.\ D {\bf 76} (2007) 014016
  [hep-ph/0702267 [HEP-PH]];
 M.~S.~Carena, A.~Menon and C.~E.~M.~Wagner,
  Phys.\ Rev.\ D {\bf 76} (2007) 035004
  [arXiv:0704.1143 [hep-ph]]; 
 J.~R.~Ellis, S.~Heinemeyer, K.~A.~Olive, A.~M.~Weber and G.~Weiglein,
  JHEP {\bf 0708} (2007) 083
  [arXiv:0706.0652 [hep-ph]];
 F.~Domingo and U.~Ellwanger,
  JHEP {\bf 0712} (2007) 090
  [arXiv:0710.3714 [hep-ph]].

\bibitem{charginolimit}
LEP2 SUSY Working Group, ALEPH, DELPHI, L3
and OPAL experiments, note LEPSUSYWG/02-05.1,
http://lepsusy.web.cern.ch/lepsusy

\bibitem{Gambin}
  M.~Misiak, H.~M.~Asatrian, K.~Bieri, M.~Czakon, A.~Czarnecki, T.~Ewerth, A.~Ferroglia and P.~Gambino {\it et al.},
  Phys.\ Rev.\ Lett.\  {\bf 98} (2007) 022002
  [hep-ph/0609232].

\bibitem{susybtosgamma}
  S.~Bertolini, F.~Borzumati, A.~Masiero and G.~Ridolfi,
  Nucl.\ Phys.\ B {\bf 353} (1991) 591.
  R.~Barbieri and G.~F.~Giudice,
  Phys.\ Lett.\ B {\bf 309} (1993) 86
  [hep-ph/9303270].
  Phys.\ Lett.\ B {\bf 322} (1994) 207
  [hep-ph/9311228].
 T.~Goto and Y.~Okada,
  Prog.\ Theor.\ Phys.\  {\bf 94} (1995) 407
  [hep-ph/9412225].


\bibitem{micromegas}
  G.~Belanger, F.~Boudjema, P.~Brun, A.~Pukhov, S.~Rosier-Lees, P.~Salati and A.~Semenov,
  Comput.\ Phys.\ Commun.\  {\bf 182} (2011) 842
  [arXiv:1004.1092 [hep-ph]].

\bibitem{btosgammaexp}
  Y.~Amhis {\it et al.}  [Heavy Flavor Averaging Group Collaboration],
  arXiv:1207.1158 [hep-ex].

\bibitem{btosgammauncertainities}
  J.~R.~Ellis, S.~Heinemeyer, K.~A.~Olive, A.~M.~Weber and G.~Weiglein,
  JHEP {\bf 0708} (2007) 083
  [arXiv:0706.0652 [hep-ph]].
\bibitem{JLMN}
Private communication with  CPsuperH author Jae Sik Lee and micrOMEGAs author Matthias Neubert.


\end{thebibliography}
\end{document}